\documentclass[12pt,preprint]{iopart}

\usepackage{epsfig}
\usepackage{graphicx}
\usepackage{amssymb}
\usepackage{iopams}
\usepackage{latexsym}
\usepackage{fontenc}
\usepackage{setspace}
\usepackage{cite}
\usepackage{soul}
\usepackage{setspace}
\usepackage[mathlines]{lineno}

 \usepackage[active]{srcltx}
\usepackage{showlabels}

\usepackage[ragged]{sidecap}

\sidecaptionvpos{table}{c}

\expandafter\let\csname equation*\endcsname\relax
\expandafter\let\csname endequation*\endcsname\relax
\usepackage{amsmath}

\usepackage[hyperindex=true,
pdftitle={Quantum Limit for Driven Linear Non-Markovian Open-Quantum-Systems},
colorlinks=true,
pagebackref=false,
citecolor=blue,
plainpages=false,
pdfpagelabels,
breaklinks=true,
linkcolor=blue,
urlcolor=blue,
]{hyperref}
\usepackage{microtype}

\newcommand{\valprom}[1]{\left<#1\right>}

\newcommand{\braket}[3]{\left<#1\right|#2\left|#3\right>}

\newcommand{\df}[1]{\mathrm{d}#1\,}
\newcommand{\dfp}[1]{\mathcal{D}#1\,}

\newcommand{\be}{\begin{equation}}
\newcommand{\ee}{\end{equation}}
\newcommand{\bes}{\begin{equation*}}
\newcommand{\ees}{\end{equation*}}
\newcommand{\bea}{\begin{eqnarray}}
\newcommand{\eea}{\end{eqnarray}}
\newcommand{\besin}{\begin{eqnarray*}}
\newcommand{\eesin}{\end{eqnarray*}}
\newcommand{\vtr}[1]{\boldsymbol{#1}}

\newcommand{\mc}[1]{\mathcal{#1}}

\begin{document}

\title{Quantum Limit for Driven Linear Non-Markovian Open-Quantum-Systems}

\author{Andr\'es F. Estrada and Leonardo A. Pach\'on}
\address{Grupo de F\'isica At\'omica y Molecular, Instituto de F\'{\i}sica,  
Facultad de Ciencias Exactas y Naturales, 
Universidad de Antioquia UdeA; Calle 70 No. 52-21, Medell\'in, Colombia.
}

\ead{andres.estrada@fisica.udea.edu.co,leonardo.pachon@udea.edu.co}

\begin{abstract}
%
The interplay between non-Markovian dynamics and driving fields in the survival
of entanglement between two non-degenerate oscillators is considered here.
Based on exact analytical results for the non-Markovian dynamics of two parametrically coupled 
non-degenerate oscillators in contact to non-identical independent thermal baths, the out-of-equilibrium 
quantum limit derived in [Phys. Rev. Lett. 105, 180501 (2010)] is generalized to the non-Markovian 
regime.
Specifically, it is shown that non-Markovian dynamics, when compared to the Markovian case, 
allow for the survival of stationary entanglement at higher temperatures, with larger coupling 
strength to the baths and at smaller driving rates.
The effect of the asymmetry of the (i) coupled oscillators, (ii) coupling strength to the baths 
at equal temperature and (iii) temperature at equal coupling strength is discussed.

\end{abstract}


\vspace{2pc}
\maketitle

\section{Introduction}
The survival of quantum features in hot environments is restricted by decoherence \cite{Zur03}. 
For quantum features such as entanglement to survive in the presence of the environment, the 
typical energy scale of the system $\hbar \omega$ must be larger than the energy scale 
$k_{\mathrm{B}}T$ associated to thermal fluctuations \cite{AW07}.
For typical nano-mechanical resonators with oscillation frequencies between 1~MHz and 
1~GHz, the quantum nature of these systems is revealed only if the temperature is well below 
10~$\mu$K and 10~mK, respectively, however, current cryostats operate at best at 15~mK \cite{OH&10}.

Recently, it was shown that the presence of driving forces relaxes the traditional criterium 
$\hbar\omega/k_{\mathrm{B}}T > 1$ discussed above  \cite{GPZ10}.
To discuss the way how this condition is relaxed, let $\gamma_{\mathrm{TB}}$ be the time scale 
associated to the non-unitary dynamics induced by the thermal bath and $\gamma_{\mathrm{p}}$ 
the pumping rate of the driving field.
In terms of these time scales, the quantum limit for driven out-of-equilibrium quantum systems 
reads
$\hbar\omega/k_{\mathrm{B}}T_{\mathrm{eff}} > 1$, where 
$T_{\mathrm{eff}} = T  \gamma_{\mathrm{TB}}/\gamma_{\mathrm{p}}$  is an effective temperature 
\cite{GPZ10,Ved10}.
For pumping rates larger than the decay rates, $\gamma_{\mathrm{TB}}/\gamma_{\mathrm{p}}<1$,
the effective temperature of the system turns out to be smaller than the thermal equilibrium temperature 
$T$. 
This allows for the survival of entanglement at higher equilibrium temperatures $T$.
For two frequency-degenerate nano-mechanical resonators of frequency $\omega = 2 \pi \times 15{\rm MHz}$
and mass $10^{-17}{\rm kg}$ with a coupling amplitude $\sim 10^{-3}m\omega^2$ and a decay rate 
$\gamma_{\mathrm{TB}}=5\times 10^{-5}\omega$, this condition implies that steady entanglement 
can be observed at tens of mK \cite{GPZ10}. 
However, entanglement might survive in the range of Kelvin if the frequency or the coupling amplitude 
can be increased by a factor 10 \cite{GPZ10}.

Very recently, a non-degenerate version of this physical model was implemented using a strongly inter-coupled 
two doubly clamped beams \cite{MO&14}. 
The coupling was mediated by an exaggerated overhang between the clamped beams.
The configuration presented in Ref.~\cite{MO&14} sustains two spectrally closely-spaced vibration modes 
at 2$\pi\times$246 and 2$\pi\times$262 kHz with quality factors of 1300 and 2200, respectively.
In combination with piezoelectric transducers, that are incorporated directly into the mechanical 
elements, it provides the key to realizing efficient parametric down-conversion \cite{MO&14}.

Moreover, there is by now experimental evidence that the non-unitary dynamics of micro-resonators 
is driven by non-Markovian dynamics \cite{GT&13}. 
This in turn suggests that a similar behaviour may be encountered at the nano-mechanical level, a 
realm where the low temperature condition already may introduce non-Markovian correlations \cite{PT&14}.
Motivated by the experimental results discussed above and by the recent interest and important role of 
non-Markovian dynamics in, e.g., biological systems \cite{IF09,NBT11,PB12c,CP&13}, quantum metrology
\cite{MBF11,CHP12}, foundations of quantum thermodynamics \cite{EG&08,PT&14} and nuclear reactions 
\cite{WS&13}, and by the intricate and delicate interplay between non-Markovian dynamics and driven 
fields in optimal control scenarios \cite{SN&11,SSA14}, the quantum limit derived in Ref.~\cite{GPZ10} 
is extended here to the case of non-Markovian dynamics.

Specially, the non-Markovian entanglement dynamics of a coupled of non-degenerate oscillators 
parametrically coupled and in contact to independent non-identical thermal baths are analytically 
solved by means of the influence functional approach of Feynman and Vernon \cite{FV63}.
Results derived here are valid for any parameter regime and allow for predicting that, when 
compared with the Markovian case, non-Markovian dynamics 
(i) decrease the time needed to generate entanglement, 
(ii) increase the temperature and the coupling-strength-to-the-environment limits at which steady 
state entanglement can be found, 
(iii) decrease the pumping rate for reaching a particular amount of entanglement and
(iv) relax the resonant driving condition.

\section{A Paradigmatic Model for Several Physical Systems}
The model described below is capable of describing a large variety 
of physical systems such as coupled  trapped ions\cite{LB&03}, coupled membranes 
or mechanical oscillators \cite{AKM14}.
Specifically, two parametrically coupled harmonic oscillators with different 
masses $m_\alpha$ and frequencies $\omega_\alpha$ are considered
\begin{equation}
\label{eq:SystmHmltnn}
H_\mathrm{S} = 
\sum_{\alpha=1}^2\frac{p_\alpha^2}{2m_\alpha}+
\frac{1}{2}m_\alpha\omega_\alpha^2q_\alpha^2
+c(t)q_1q_2,
\end{equation}
where $c(t)$ is an arbitrary coupling function between the oscillators.
This time dependence is critical for the generation and maintenance of entanglement
\cite{GPZ10,SN&11}.

To describe the interaction with their surroundings (dissipative and decoherencing effects) 
in a rigorously way,  the system-bath model \cite{Wei08,BP02,FV63,CL83,GSI88} in the 
context of the Caldeira-Leggett model is considered here.
To avoid extra correlations between the oscillators from sharing a common thermal bath 
\cite{PR08,RR13}, the two oscillators are coupled to independent thermal baths with 
different power noise [different spectral density $J(\omega)$ and different temperature $T$], 
see below.
The Hamiltonian of the baths $\hat{H}_{\mathrm{B}}$, including the interaction with the system 
of interest $\hat{H}_{\mathrm{I}}$, is then given by
\begin{equation}
\label{eq:BathSystmHmltnn}
H_\mathrm{B} + H_\mathrm{I}=\sum_{\alpha=1}^2\sum_{k=1}^N
\frac{p_{\alpha,k}^2}{2m_{\alpha,k}}
+\frac{1}{2}m_{\alpha,k}\omega_{\alpha,k}^2 
\left(
q_{\alpha,k} - \frac{c_{\alpha,k}}{m_{\alpha,k}\omega_{\alpha,k}^2} q_{\alpha}
\right)^2,
\end{equation}
where the coefficients $c_{\alpha,k}$ are the coupling constants among the oscillators of 
the system of interest and each of the modes of their own thermal baths. 
It can be seen that the interaction is bilinear in the position operators of the systems and 
the thermal baths.
This implies considering only the linear response of the thermal baths 
to the influence of the system. 
This  consideration is valid only for the case of geometrically macroscopic thermal baths, 
which leads to a weak interaction among the oscillators in the system and 
each one of the oscillators comprising the baths\cite{Wei08,CL83}. 
Note that in the interaction Hamiltonian (\ref{eq:BathSystmHmltnn}), there are two terms 
that only depends on $q_{\alpha}$ and on the coupling constants to the thermal baths. 
Those terms are included to compensate the renormalization of the harmonic potentials 
in the system of interest by the presence of the thermal baths \cite{Wei08,CL83,Ing02}. 
By considering these terms, it is ensured that the minimum of the global Hamiltonian with 
respect to the system-of-interest coordinates is determined only by the potentials in the 
system \cite{Ing02}.

\subsection{Analytic Exact System Dynamics}
The dynamics of the coupled oscillators are solved by means of the influence functional 
approach\cite{FV63}. Details can be found in \ref{app:InfncFnctnl} below. 
In solving for the dynamics, the initial density operator of the oscillator couple 
$\hat{\rho}_{\mathrm{S}}$ and their thermal baths $\hat{\rho}_{\mathrm{TB}_1}$ and 
$\hat{\rho}_{\mathrm{TB}_2}$ are assumed to factorize, i.e., 
\begin{align}
\label{eq-17}
\hat{\rho}(0) = \hat{\rho}_\mathrm{S}(0)\otimes \hat{\rho}_{\mathrm{TB}_1}(0)
\otimes \hat{\rho}_{\mathrm{TB}_2}(0),
\end{align}
that is, it is supposed that at time $t=0$ there are no initial correlations between the 
subsystems of the overall system. 
However, this assumption may not always valid because in many applications both, the 
degrees of freedom of the system of interest and the environment to which it is attached 
form part of the same system.
Thus, it is possible that the initial correlations are not available to the experimentalist \cite{GSI88}. 
Although, initial conditions may be relevant in the generation of new control strategies in 
open quantum systems \cite{PB13,PYB13,PB13c,PB14a}, they are not considered below
for the sake of  concreteness.

According to the influence functional approach, at time $t$ the matrix elements 
$\braket{q_{1+}'',q_{2+}''}{\hat{\rho}_\mathrm{S}(t)}{q_{1-}'',q_{2-}''}$ of the density operator 
$\hat{\rho}_{\mathrm{S}}(t)$ are given by
\begin{align}
\label{eq-19}
&\braket{\mathbf{q}_{+}''}{\hat{\rho}_\mathrm{S}(t)}{\mathbf{q}_{-}''} =
\int_{-\infty}^\infty \mathrm{d}^2 q_{+}' \mathrm{d}^2 q_{-}'
J(\mathbf{q}_{+}'',\mathbf{q}_{-}'',t; \mathbf{q}_{+}',\mathbf{q}_{-}',0)
\braket{\mathbf{q}_{+}'}{\hat{\rho}_\mathrm{S}(t)}{\mathbf{q}_{-}'},
\end{align}
where $\mathbf{q}_{\pm} = (q_{1\pm},q_{2\pm})$ and the propagating function of the reduced 
density matrix $J(\mathbf{q}_{+}'',\mathbf{q}_{-}'',t; \mathbf{q}_{+}',\mathbf{q}_{-}',0) $
reads
\begin{align}
\label{eq-Jpropfunct}
&J(\mathbf{q}_{+}'',\mathbf{q}_{-}'',t; \mathbf{q}_{+}',\mathbf{q}_{-}',0) =
\int \mathcal{D}^2 q_{+} \mathcal{D}^2 q_{-}
\exp\biggl\{\frac{\mathrm{i}}{\hbar}
\left(S_\mathrm{S}[\mathbf{q}_{+}]-S_\mathrm{S}[\mathbf{q}_{-}]\right)\biggr\}
\mc{F}[\mathbf{q}_{+},\mathbf{q}_{-}].
\end{align}
$\mc{F}[\mathbf{q}_{+},\mathbf{q}_{-}]$ denotes the influence functional\cite{FV63,FH05} 
and is given by [see \ref{app:InfncFnctnl}]
\begin{equation}
\label{ap-19}
\begin{split}
&\mc{F}[\mathbf{q}_{+},\mathbf{q}_{-}]
 =\prod_{\alpha=1}^2 \exp\biggl(-\frac{\mathrm{i}}{\hbar}\frac{m_\alpha}{2}
\biggl\{(q_{\alpha +}'+q_{\alpha -}')\int_0^t\df{s}\gamma_\alpha(s)
\left[q_{\alpha +}(s)-q_{\alpha-}(s)\right]\biggr.\biggr.  
\\
&\phantom{f(a)} + \biggl.\biggl.\int_0^t\df{s}
\int_0^s\df{u}\gamma_\alpha(s-u)\left[\dot{q}_{\alpha +}(u)+
\dot{q}_{\alpha-}(u)\right]
\left[q_{\alpha +}(s)-q_{\alpha -}(u)\right]\biggr\}\biggr) 
\\
&\phantom{f(a)} \times \exp\biggl\{-\frac{1}{\hbar}\int_0^t\df{s}
\int_0^s\df{u}K_\alpha(u-s)\left[q_{\alpha +}(s)
- q_{\alpha-}(s)\right]\left[q_{\alpha +}(u)-q_{\alpha -}(u)\right]\biggr\}.
\end{split}
\end{equation}
The dissipation kernel $\gamma_\alpha(s)$ and noise kernel $K_\alpha(s)$ are 
defined in terms of the spectral density $J_\alpha(\omega)$ as
\begin{align}
\label{eq-krnlgamma} 
\gamma_\alpha(s) &= 
\frac{2}{m_\alpha}\int_0^\infty\frac{\df{\omega_\alpha}}{\pi}J_\alpha(\omega_\alpha)\cos(\omega_\alpha s), 
\\
\label{eq-krnlK}
K_\alpha(s) &= \int_0^\infty\frac{\df{\omega_\alpha}}{\pi}J_\alpha(\omega_{\alpha})
\coth\left(\frac{\hbar\beta_\alpha\omega_\alpha}{2}\right)\cos(\omega_\alpha s), 
\end{align}
where $\beta_\alpha=1/k_\mathrm{B}T_\alpha$, being $k_\mathrm{B}$ the Boltzmann 
constant and $T_\alpha$ the temperature of each of the thermal baths.
The spectral density $J_{\alpha}(\omega_\alpha) = \pi\sum_{k=1}^N 
\frac{c_{\alpha,k}^2}{2m_{\alpha,k}\omega_{\alpha,k}}\delta(\omega_\alpha-\omega_{\alpha,k})$
comprises all the information of the bath that is needed to account for its influence on the system.

In deriving an exact closed expression for the propagating function, it is necessary 
to evaluate the four-fold path integral in Eq.~(\ref{eq-Jpropfunct}).
The exact path integration is performed by taking advantage of the linearity of the system 
and by using standard techniques \cite{Wei08,Ing02}. 
Details on the derivation can be found in \ref{app:InfncFnctnl}.
The propagating function is conveniently written as
\begin{equation}
\label{eq:FnlPrpFcntin}
\begin{split}
J(\mathbf{Q}'',\mathbf{q}'',t; \mathbf{Q}', \mathbf{q}',0) &= 
 \frac{1}{N(t)} \exp\biggl\{\frac{\mathrm{i}}{\hbar}\sum_{\alpha=1}^2\left[q_\alpha''\dot{Q}_\alpha(t)
-q_\alpha'\dot{Q}_\alpha(0)\right]\biggr. \\
\biggl.&-\frac{1}{\hbar}\int_0^t\df{s}\int_0^s \df{u}\sum_{\alpha=1}^2K_\alpha(u-s)q_\alpha(s)q_\alpha(u)\biggr\}.
\end{split}
\end{equation}
with $Q_\alpha=\frac{1}{2}(q_{\alpha +} + q_{\alpha -})$, 
$q_\alpha = q_{\alpha +} - q_{\alpha -}$  and  $N(t)$ is a normalization factor that can be 
determined from the conservation of the normalization of the density matrix, 
$\mathrm{tr}\hat{\rho}_{\mathrm{S}}=1$,
\begin{equation}
\label{equ:Noft}
N(t)= \pi^2\hbar^2/\left| A_{16}(t)A_{38}(t)-A_{18}(t)A_{36}(t)\right|,
\end{equation}
the matrix elements $A_{ij}$ are defined in Eq.~(\ref{ap-35}).
The new coordinates $Q_\alpha$ and $q_\alpha$ satisfy the following equation 
of motion
\begin{align}
\label{eq:ExtrmSltn}
\begin{split}
\ddot{Q}_{1,2}(s)+\omega_{1,2}^2 Q_{1,2}(s) + 
\frac{c(s)}{m_{1,2}}Q_{2,1}(s)+\frac{\mathrm{d}}{\mathrm{d}s}
\int_0^s \df{u} \gamma_{1,2}(s-u)Q_{1,2}(u) &= 0, 
\\
\ddot{q}_{1,2}(s)+\omega_{1,2}^2 q_{1,2}(s) + 
\frac{c(s)}{m_{1,2}} q_{2,1}(s) - \frac{\mathrm{d}}{\mathrm{d}s} 
\int_s^t \df{u} \gamma_{1,2}(u-s)q_{1,2}(u) &=0,
\end{split}
\end{align}
with the boundary conditions for $Q_\alpha(0)=Q_\alpha '$, $Q_\alpha(t)=Q_\alpha ''$, 
$q_\alpha(0)=q_\alpha '$, $q_\alpha(t)=q_\alpha ''$.

In the original frequency-and-mass degenerate formulation of the driving-assisted-high-temperature-entanglement 
scenario\cite{GPZ10}, the spectral densities $ J_1(\omega) = J_2(\omega) = J(\omega)$ were 
taken in the Ohmic form $J(\omega)= m \gamma\omega$ with $\gamma$ denoting the standard
coupling to the environment constant.
This leads to local-in-time dissipation in Eqs.~(\ref{eq:ExtrmSltn}) provided by the fact that 
$\gamma_{1,2}(s)=2\gamma\delta(s)$, and induces larger decay rates for the loss of entanglement
(see below). 
To overcome this and to have a more general and complete characterization of the dynamics
that allows for a closer description of experimental realization \cite{GT&13,MO&14}, 
the influence of the bath on the system is characterized here by the spectral density
\begin{equation}
\label{eq:Jomega}
J_\alpha(\omega)=m_\alpha\gamma_\alpha\omega \frac{\Omega_\alpha^2}{\omega ^2+\Omega_\alpha^2},
\end{equation}
where $\Omega_\alpha$ denotes a finite cut-off frequency.
By replacing the last expression in Eq.~(\ref{eq-krnlgamma}), this spectral density generates
$
\gamma_\alpha(s)=\gamma_\alpha\Omega_\alpha\exp\left(-\Omega_\alpha\left|s\right|\right),
$
which leads to memory effects in the dissipation for times $s<\tau_\alpha=\Omega_\alpha^{-1}$ 
in the equations of motion (\ref{eq:ExtrmSltn}).
For this spectral density
$
K_\alpha(s)=m_\alpha\gamma_\alpha\Omega_\alpha^2(\hbar\beta_\alpha)^{-1}
\sum_{n=-\infty}^\infty \left[\Omega_\alpha\exp\left(-\Omega_\alpha|s|\right)
-|\nu_{\alpha,n}|\exp\left(-|\nu_{\alpha,n}||s|\right)\right] \left[\Omega_\alpha^2-\nu_{\alpha,n}^2\right]^{-1},
$
being $\nu_{\alpha,n}=2\pi n (\hbar\beta_\alpha)^{-1}$ Matsubara's frequencies\cite{Wei08,BP02,Ing02}. 

For the degenerate Ohmic situation in Ref.~{\cite{GPZ10}}, the solution to 
Eqs.~(\ref{eq:ExtrmSltn}) can be expressed in terms of the solutions of Mathieu's oscillator 
(see also Ref.~\cite{ZH95}).
The non-Ohmic character of the spectral density in Eq.~(\ref{eq:Jomega}) prevents here the 
formulation of the solution of Eqs.~(\ref{eq:ExtrmSltn}) in a similar fashion.
However, their linear character allows for expressing the formal solution
in the form
\begin{align}
\label{eq:SltnsExtrmCndtns}
\begin{split}
Q_1(t,s) &= U_1(t,s) Q_1' +U_2(t,s) Q_1''+ U_3(t,s)Q_2'+U_4(t,s)Q_2'', \\
Q_2(t,s) &= V_1(t,s)Q_2'+V_2(t,s)Q_2''+V_3(t,s)Q_1'+V_4(t,s)Q_1'', \\
q_1(t,s) &= u_1(t,s)q_1'+u_2(t,s)q_1''+u_3(t,s)q_2'+u_4(t,s)q_2'', \\
q_2(t,s) &= v_1(t,s)q_2'+v_2(t,s)q_2''+v_3(t,s)q_1'+v_4(t,s)q_1'',
\end{split}
\end{align}
where this set of sixteen auxiliary functions $\{U_i,V_i,u_i,v_i\}$ is obtained by numerical
integration of the associated set of second order non-local-in-time differential equations 
that arises for $\{U_i,V_i,u_i,v_i\}$  after plugging (\ref{eq:SltnsExtrmCndtns}) into Eq.~(\ref{eq:ExtrmSltn}).
Because of the time-reversed character of the limits in the integral contribution to the equation 
of motion of coordinate $q_{\alpha}$ in Eq.~(\ref{eq:ExtrmSltn}), special care must be exercised 
in the numerical integration.
Note that a direct numerical integration of (\ref{eq:ExtrmSltn}) would not allow for deriving 
the analytic result for the propagating function in Eq.~(\ref{eq:FnlPrpFcntin}).

\section{Entanglement Quantification and Covariance Matrix Elements}
Due to the linearity of the system's Hamiltonian (\ref{eq:SystmHmltnn}) and 
the Gaussian character of the propagating function of the reduced density matrix 
in Eq.~(\ref{eq:FnlPrpFcntin}), every initial Gaussian state evolves into another 
Gaussian state.
For the present kind of bipartite system of continuous variables in a Gaussian state, entanglement 
can be easily quantified in terms of the logarithmic negativity \cite{VW02}.
This quantity gives a characterization of the amount of entanglement which can be 
distilled into singlets. 

To quantify entanglement in this case, Gaussian continuous variable states, only 
the covariance matrix $\sigma$ is needed. 
It reads
%
$\sigma_{ij}=\frac{1}{2}\valprom{\hat{\xi}_i \hat{\xi}_j+\hat{\xi}_j \hat{\xi}_i} - 
\valprom{\hat{\xi}_i} \valprom{\hat{\xi}_j}$,
where $\hat{\boldsymbol{\xi}}=(\hat{q}_1,\hat{q}_2,\hat{p}_1,\hat{p}_2)$ and 
$\hat{q}_1$, $\hat{q}_2$, $\hat{p}_1$, $\hat{p}_2$ are the position and 
momentum operators of the oscillators in the system of interest characterized by the Hamiltonian
(\ref{eq:SystmHmltnn}). 

\subsection{Entanglement Quantification}
\label{subsub:EntangQuant}
The logarithmic negativity is defined as
$
E_{\mathrm{N}}=-\frac{1}{2}\sum_{i=1}^4\log_2[\min(1,2|l_i|)],
$ 
where $l_i$'s are the symplectic eigenvalues of the covariance matrix. 
They are the normal eigenvalues of the matrix $-\mathrm{i}\Sigma\sigma$, with
$
\Sigma=\left(\begin{array}{cc}
0&\mathsf{I}_2\\
-\mathsf{I}_2&0
\end{array}\right)
$
the symplectic matrix and $\mathsf{I}_2$ the identity matrix of dimension 2.
For separable states, $\hat{\rho}_{\mathrm{S}}=\sum_i p_i \hat{\rho}_1^{(i)}\otimes \hat{\rho}_2^{(i)}$,
the logarithmic negativity of the system is zero and each oscillator can be described independently. 
For continuous variable systems, entanglement has as upper limit the maximally entangled EPR 
wave-function with $E_{\mathrm{N}}\to\infty$.
Hence, the amount of entanglement measured by the logarithmic negativity is unbounded from 
above.

\subsection{Covariance matrix elements}
To calculate any mean value of any operator associated with the observables of one of the
oscillators in the system of interest, it is necessary to find the reduced density matrix associated 
to that oscillator.
This is obtained by tracing out the reduced density matrix over the coordinates of the other 
oscillator, i.e., 
$\hat{\rho}_{\mathrm{S}_{1,2}} = \mathrm{tr}_{\mathrm{S}_{2,1}} \hat{\rho}_{\mathrm{S}}$.
For instance, the matrix elements of the reduced density operator associated with the first oscillator 
are given by
\begin{equation}
\label{eq-42}
\braket{Q_1''}{\hat{\rho}_\mathrm{S_1}(t)}{q_1''}=
\int_{-\infty}^\infty\df{Q_2''} \braket{Q_1'' ,Q_2''}{\hat{\rho}_\mathrm{S}(t)}{q_1'',q_2''=0},
\end{equation}
while for the second oscillator
$
\braket{Q_2''}{\hat{\rho}_\mathrm{S_2}(t)}{q_2''}=
\int_{-\infty}^\infty\df{Q_1''} \braket{Q_1'' ,Q_2''}{\hat{\rho}_\mathrm{S}(t)}{q_1''=0,q_2''}.
$

To find the first and second moments of each oscillator, it is necessary to perform 
the following integrals:
\begin{equation}
\label{eq:frstscndmmnts}
\begin{split}
\valprom{\hat{q}_\alpha}(t) &=
\int_{-\infty}^\infty\df{Q_\alpha''}Q_\alpha''\, 
\braket{Q_\alpha''}{\hat{\rho}_\mathrm{S_\alpha}(t)}{q_\alpha''=0},  
 \\ 
\valprom{\hat{p}_\alpha}(t) &= 
-\mathrm{i}\hbar\int_{-\infty}^\infty\df{Q_\alpha''}\frac{\mathrm{d}}{\mathrm{d}q_\alpha''}
\biggl.\braket{Q_\alpha''}{\hat{\rho}_\mathrm{S_\alpha}(t)}{q_\alpha''}\biggr|_{q_\alpha''=0}, 
 \\
\valprom{\hat{q}_\alpha^2}(t) &=
 \int_{-\infty}^\infty\df{Q_\alpha''}Q_\alpha''^2\, 
 \braket{Q_\alpha''}{\hat{\rho}_\mathrm{S_\alpha}(t)}{q_\alpha''=0}, 
 \\ 
\valprom{\hat{p}_\alpha^2}(t) &= 
-\hbar^2\int_{-\infty}^\infty\df{Q_\alpha''}\frac{\mathrm{d}^2}{\mathrm{d}q_\alpha''^2}
\biggl.\braket{Q_\alpha''}{\hat{\rho}_\mathrm{S_\alpha}(t)}{q_\alpha''}\biggr|_{q_\alpha''=0}, 
 \\
\frac{1}{2}\valprom{\hat{q}_\alpha \hat{p}_\alpha + \hat{p}_\alpha \hat{q}_\alpha}(t) &= 
-\mathrm{i}\hbar\int_{-\infty}^\infty\df{Q_\alpha''}
Q_\alpha''\frac{\mathrm{d}}{\mathrm{d}q_\alpha''}
\biggl.\braket{Q_\alpha''}{\hat{\rho}_\mathrm{S_\alpha}(t)}{q_\alpha''}\biggr|_{q_\alpha''=0}.
\end{split}
\end{equation}
To find the covariances between the position and/or momentum operators between 
the oscillators, it is necessary to integrate out
\begin{equation}
\label{eq:scndmmntsmxd}
\begin{split}
\valprom{\hat{q}_1 \hat{q}_2}(t) &= 
\int_{-\infty}^\infty\df{Q_1''}\df{Q_2''}Q_1''\,Q_2''\,
\braket{Q_1'' ,Q_2''}{\hat{\rho}_\mathrm{S}(t)}{q_1''=0,q_2''=0}, 
 \\
\valprom{\hat{q}_1 \hat{p}_2}(t) &= 
-\mathrm{i}\hbar\int_{-\infty}^\infty\df{Q_1''}\df{Q_2''}Q_1''
\frac{\mathrm{d}}{\mathrm{d}q_2''}\biggl.
\braket{Q_1'' ,Q_2''}{\hat{\rho}_\mathrm{S}(t)}{q_1''=0,q_2''}\biggr|_{q_2''=0}, 
 \\
\valprom{\hat{q}_2 \hat{p}_1}(t) &= 
-\mathrm{i}\hbar\int_{-\infty}^\infty\df{Q_1''}\df{Q_2''}Q_2''
\frac{\mathrm{d}}{\mathrm{d}q_1''}
\biggl.\braket{Q_1'' ,Q_2''}{\hat{\rho}_\mathrm{S}(t)}{q_1'',q_2''=0}\biggr|_{q_1''=0}, 
 \\
\valprom{\hat{p}_1 \hat{p}_2}(t) &= 
-\hbar^2\int_{-\infty}^\infty\df{Q_1''}\df{Q_2''}\frac{\mathrm{d}^2}{\mathrm{d}q_1''\,\mathrm{d}q_2''}
\biggl.\braket{Q_1'' ,Q_2''}{\hat{\rho}_\mathrm{S}(t)}{q_1'',q_2''}\biggr|_{q_1''=0,q_2''=0}.
\end{split}
\end{equation}
For the case when no initial correlations exist between the oscillators in the system of interest,
namely, $\hat{\rho}_\mathrm{S}(0)=\hat{\rho}_{\mathrm{S}_1}(0) \otimes \hat{\rho}_{\mathrm{S}_2}(0)$, 
the exact analytic expression for the first and second moments as functions of the set of auxiliary 
functions $\{U_i,V_i,u_i,v_i\}$ can be found at 
\href{http://gfam.udea.edu.co/~lpachon/scripts/oqsystems>}{http://gfam.udea.edu.co/~lpachon/scripts/oqsystems}.
Note that the expressions for the first and second moments (\ref{eq:frstscndmmnts}) and the 
mixed moments (\ref{eq:scndmmntsmxd}) were calculated for arbitrary driving forces $c(t)$.
This allows for using expressions (\ref{eq:frstscndmmnts}) and (\ref{eq:scndmmntsmxd}) in the
context, e.g., of optimal control of sideband cooling of nanomechanical resonators \cite{TEP14}.

\subsection{Initial Gaussian states for the simulations}
To obtain specific results, both oscillators are assumed in a general Gaussian state,
and therefore, the initial density matrix for each oscillator can be expressed in terms of the 
coordinates $Q_\alpha$ and $q_\alpha$ as
\begin{align}
\label{eq:initialstate}
\rho_{\alpha}(Q_\alpha',q_\alpha',0) &= 
\frac{1}{\sqrt{2\pi \sigma_q^{(\alpha)}}}\exp\Biggl\{-\frac{1}{2\sigma_{qq}^{(\alpha)}}
\left(Q_\alpha'-\valprom{q_{\alpha}}\right)^2
-\frac{1}{2\hbar^2}\left(\sigma_{pp}^{(\alpha)}-\frac{\sigma_{qp}^{2\,(\alpha)}}{\sigma_{qq}^{(\alpha)}}\right)q_\alpha'\Biggr. 
\notag \\
&+ \Biggl.\frac{\mathrm{i}}{\hbar}\left[\valprom{p_{\alpha}}+\frac{\sigma_{qp}^{(\alpha)}}{\sigma_{qq}^{(\alpha)}}
\left(Q_\alpha'-\valprom{q_{\alpha}}\right)\right]q_\alpha'\Biggr\},
\end{align}
where $\valprom{q_{\alpha}}$, $\valprom{p_{\alpha}}$, $\sigma_{qq}^{(\alpha)}$, 
$\sigma_{pp}^{(\alpha)}$ and $\sigma_{qp}^{(\alpha)}$ are the first moments of position and momentum, 
and the variance of the position, momentum and position-momentum, respectively, 
for the $\alpha$-th oscillator.

\section{Entanglement Dynamics for Symmetric Thermal Baths}
To characterize the influence of the non-Markovian dynamics in reaching a different quantum 
limit, symmetric thermal bath are considered at this point, i.e., $\gamma_1 = \gamma_2 = \gamma$,  
$\Omega_1 = \Omega_2 = \Omega$ and  $T_1 = T_2 = T$.
Moreover, to compare with results in Ref.~\cite{GPZ10}, $c(t)$ is chosen here as 
\begin{equation}
\label{eq-chrntpmpng}
c(t)=c_1\cos(\omega_{\mathrm{d}} t),
\end{equation}
where $\omega_{\mathrm{d}}$ denotes the frequency of the driving field and $c_1$
its constant amplitude. 
Although results presented below are particular to the spectral density in Eq.~($\ref{eq:Jomega}$),
they encompass the most basic feature of non-Markovian dynamics, namely, a non-flat power 
noise \cite{PT&14}.

Under the conditions stablished above, and for degenerate oscillators $\omega_1 = \omega_2 = \omega$
and $m_1 = m_2 = m$, entanglement is quantify below by means of the logarithmic negativity 
introduced in Sec.~\ref{subsub:EntangQuant}.
The undriven non-Markovian dynamics for this degenerate case was 
previously analyzed in Ref.~\cite{MOP07} and it was found that when $\omega$ falls inside 
the spectral density (non-Markovian case), entanglement persists for a longer time than in 
the Markovian case.
For the present driven case, the dynamic features provided by Markovian [$\Omega_\alpha = 20 \omega$ in 
Eq.~(\ref{eq:Jomega})] and non-Markovian [$\Omega_\alpha = \omega$ in Eq.~(\ref{eq:Jomega})] 
are presented in Figs.~\ref{fig:equal_tempdep} and \ref{fig:equal_gammadep} for a variety 
of temperatures $T$ and coupling constants $\gamma$, respectively.
The main feature depicted by Figs.~\ref{fig:equal_tempdep} and \ref{fig:equal_gammadep}
is that if a given amount of entanglement is reached under Markovian dynamics, the same 
amount can be found under non-Markovian dynamics, but for temperatures or coupling 
constants as higher as the double of the value of the parameters under Markovian 
dynamics.
In Sec.~\ref{subsub:non-MarkQntmLmt}, a quantitative discussion on this regards is provided.

\subsection{Entanglement dynamics as a function of the bath temperature $T$}
%
Figure~\ref{fig:equal_gammadep} shows the time evolution of the logarithmic negativity for a variety 
of coupling temperatures.
Dark thick curves correspond to the non-Markovian case [$\Omega = \omega$ in 
Eq.~(\ref{eq:Jomega})] and light thin curves correspond to the Markovian case 
[$\Omega = 20 \omega$ in Eq.~(\ref{eq:Jomega})].
Form there, it is clear that for the particular functional form of $J(\omega)$ in Eq.~(\ref{eq:Jomega}),
non-Markovian dynamics are able to support the same amount of steady state entanglement at 
twice the temperature than the corresponding Markovian case.
\begin{SCfigure}[][h!]
\centering
\caption{Entanglement dynamics as a function of the temperature $T$ for Markovian, 
$\Omega=20\omega$ (M), and non-Markvovian dynamics,  $\Omega=\omega$  (nM).
Parameter values are $\gamma=10^{-3}\omega$, $c_1 = 0.2 m \omega^2$, 
and $\omega_{\mathrm{d}} = 2 \omega$.}
\includegraphics[width=0.6\columnwidth]{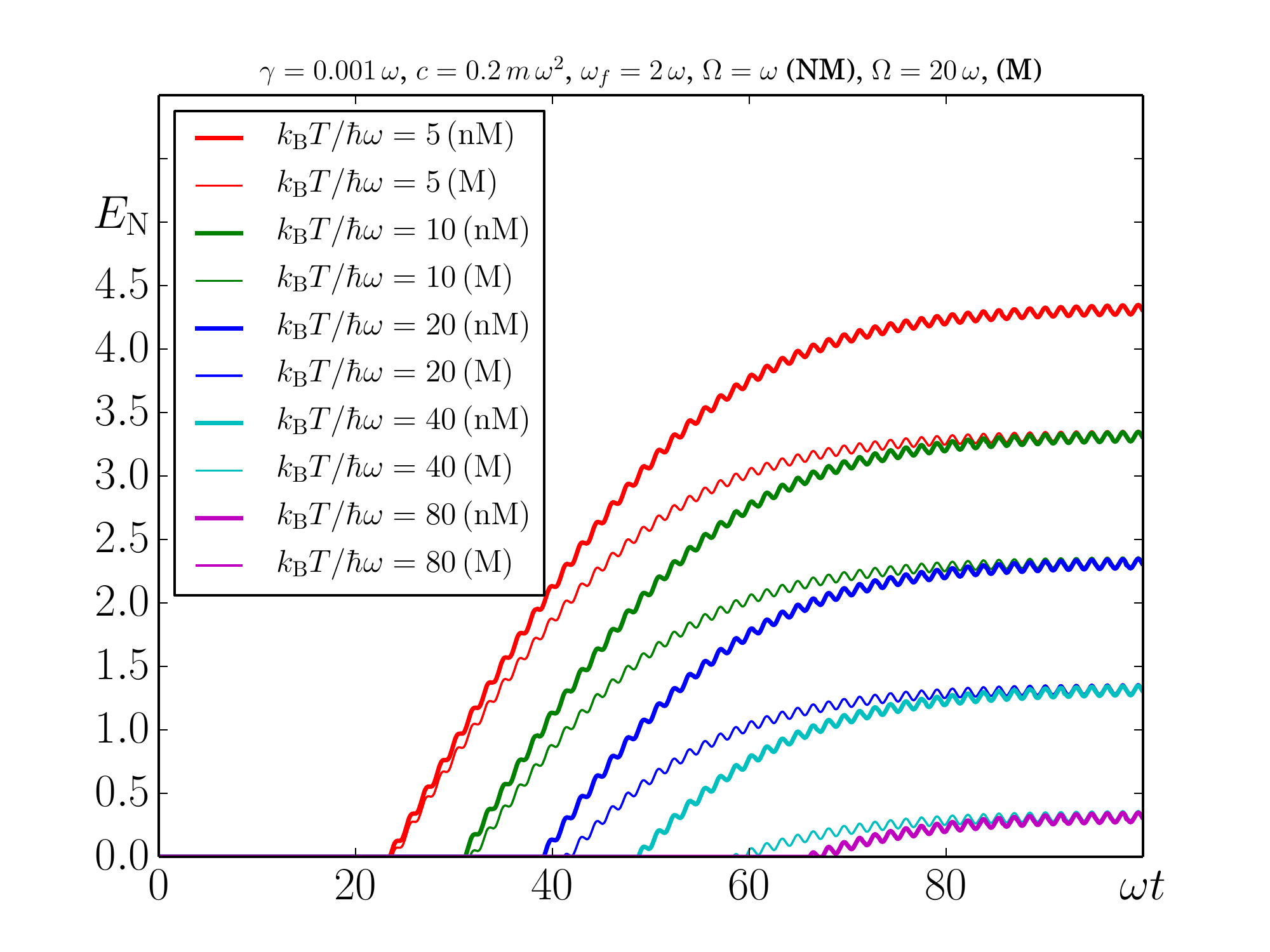}
\label{fig:equal_tempdep}
\end{SCfigure}

An additional dynamic feature present in Figs.~\ref{fig:equal_gammadep} and 
\ref{fig:equal_tempdep} is that entanglement is generated at shorter times under non-Markovian 
dynamics than in the Markovian case.
Since the rate of the incoherent processes decreases by non-Markovian dynamics 
(see Sec.~\ref{subsub:non-MarkQntmLmt}), then it is natural to expect that the driving force
needs to preform less work to squeeze the normal modes of the oscillators when the dynamics 
are non-Markovian.

\subsection{Entanglement dynamics as a function of the bath coupling constant $\gamma$}
\begin{SCfigure}[][h!]
\centering
\caption{Entanglement dynamics as a function of the constant coupling  
$\gamma$ for Markovian, $\Omega=20\omega$  (M), and non-Markvovian dynamics,  
$\Omega=\omega$  (nM).
Parameter values are $\gamma=10^{-3}\omega$, $k_{\mathrm{B}}T = 5 \hbar \omega$, 
$c_1 = 0.2 m \omega^2$, and $\omega_{\mathrm{d}} = 2 \omega$.}
\includegraphics[width=0.6\columnwidth]{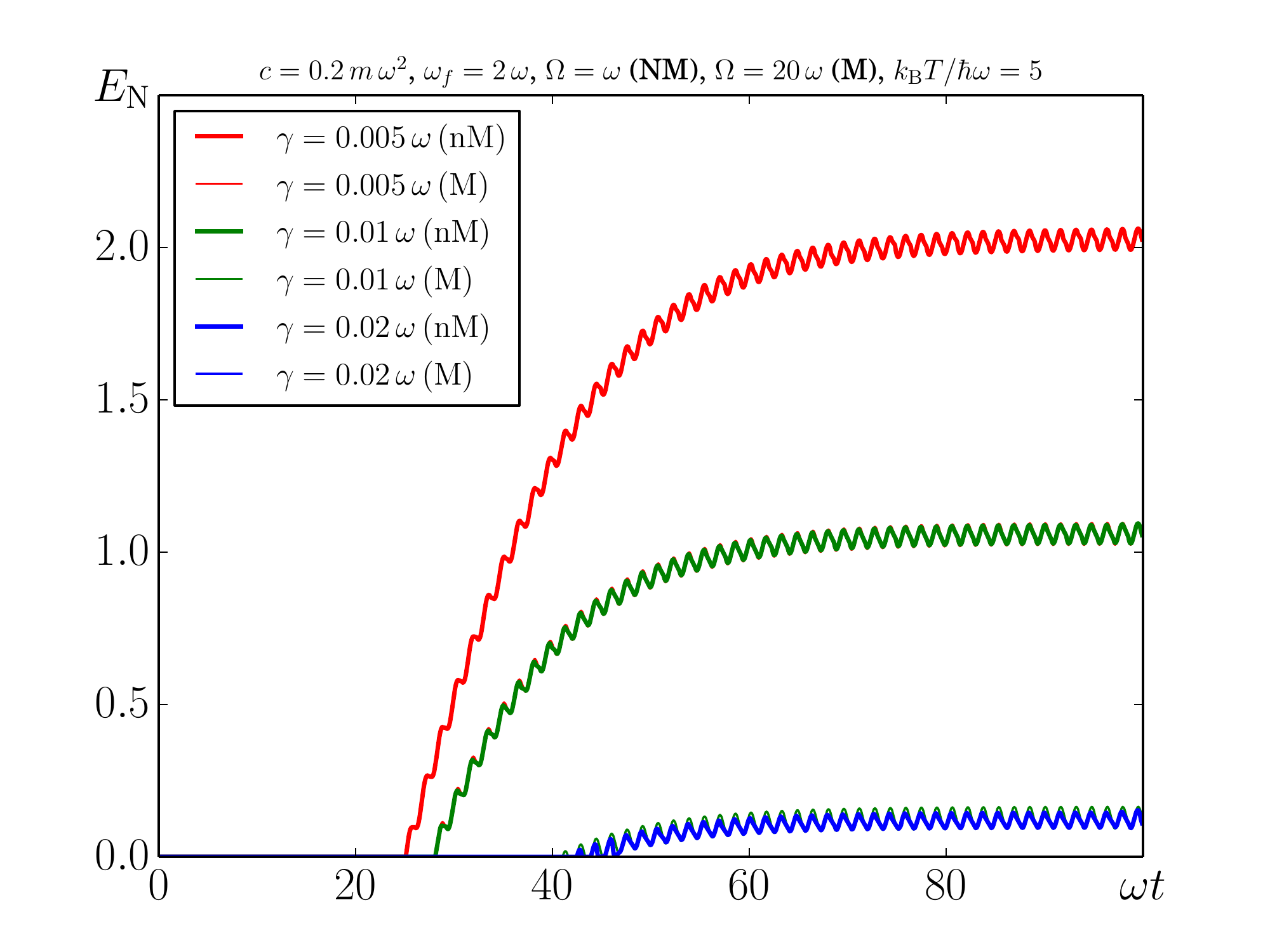}
\label{fig:equal_gammadep}
\end{SCfigure}
Figure~\ref{fig:equal_gammadep} depictes the time dynamics of entanglement for a variety 
of coupling rates $\gamma$ at fixed temperature.
In analogy to the case discussed in Fig.~\ref{fig:equal_tempdep}, non-Markovian dynamics 
are able to support the same amount of steady state entanglement at twice the coupling rate
$\gamma$ than the corresponding Markovian case.
Because of the non-Markovian character of the dynamics, simulations over several periods of 
the driving force are very expensive on computational terms, the amplitude strength $c_1$ 
takes a rather large value ($c_1 = 0.2 m \omega^2$) so that the generation of entanglement 
occurs after a few periods of driving.
However,  the effects discussed above are clearly present for smaller values of the  amplitude 
strength $c_1$ (see Sec.~\ref{subsub:non-MarkQntmLmt}).

\subsection{Entanglement dynamics as a function of the initial state}
\begin{SCfigure}[][h!]
\centering
\caption{Non-Markovian entanglement dynamics for a variety  of different initial states. 
Parameter values are $k_{\mathrm{B}}T= 10 \hbar \omega$, $\gamma=10^{-3}\omega$, 
$c_1 = 0.2 m \omega^2$, and $\omega_{\mathrm{d}} = 2 \omega$.}
\includegraphics[width=0.6\columnwidth]{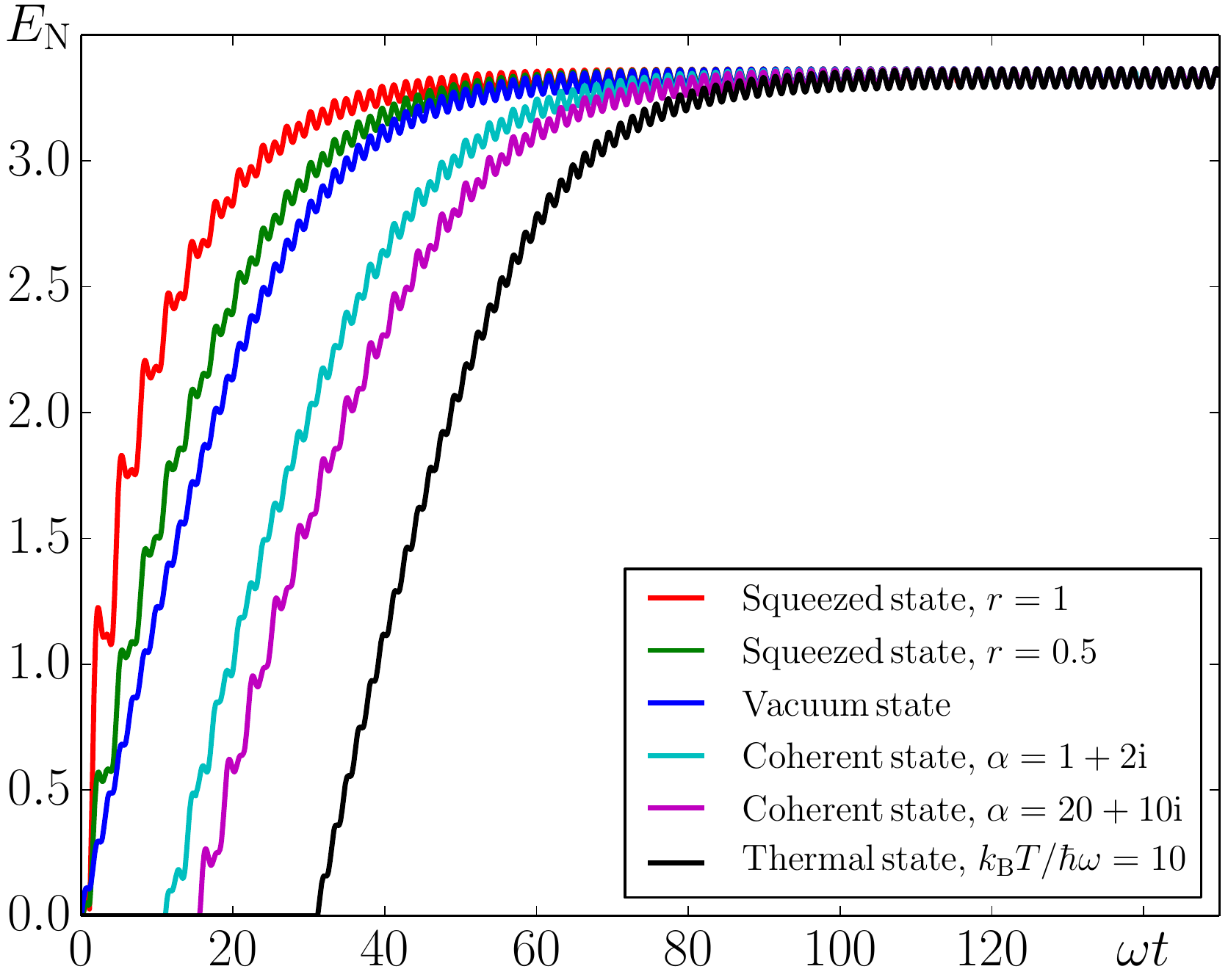}
\label{fig:initiastatepdep}
\end{SCfigure}

One of the most attractive features of the generation of entanglement by driving forces in 
the presence of non-unitary dynamics is that the system reaches the same amount of stationary 
entanglement independently of the its initial state \cite{GPZ10}. 
Despite of the non-Markovian dynamics and its associated dependence of the history of the 
system evolution, this feature remains present here. 
Figure~\ref{fig:initiastatepdep} shows the time evolution of the logarithmic negativity for a variety 
initial states given by Eq.~(\ref{eq:initialstate}).
There, it is clear that all of them reach the same amount of stationary entanglement.

\subsection{Non-Markovian Quantum Limit}
\label{subsub:non-MarkQntmLmt}
If the characteristic frequency of a given system is denoted by $\omega$, the survival of quantum 
features such as entanglement can be predicted for the parameter relations satisfying the condition 
$S(\omega)< \omega \mu$, being $S(\omega) = \frac{1}{2m}J(\omega)\coth(\frac{1}{2}\hbar \omega \beta)$ 
the power noise and $\mu$ the pumping rate.
Specifically, at high temperate $\hbar \omega \beta \ll 1$, entanglement survives in the steady 
state if 
\begin{equation}
\label{equ:nMqlimit}
k_{\mathrm{B}}TJ(\omega)/m < \hbar \omega^2 \mu.
\end{equation}
Because of the various time/energy scales involved in this non-trivial out-of-equilibrium situation, 
the impact of non-Markovian dynamics can understood in manifold ways, specifically, it can be 
\emph{effectively} ascribed to each time/energy scale independently.
In doing so, the case of the spectral density in Eq.~(\ref{eq:Jomega}) is discussed below and 
three non-Markovianly scaled parameters, $T_{\mathrm{nM}} $, $\gamma_{\mathrm{nM}} $ and 
$\mu_{\mathrm{nM}} $, are introduced.

After plugging the spectral density (\ref{eq:Jomega}) in the non-Markovian quantum limit given by
Eq.~(\ref{equ:nMqlimit}), a new effective temperature can be defined
\begin{equation}
T_{\mathrm{nM}} = T \left( 1 - \frac{1}{1+\Omega^2/\omega^2} \right),
\end{equation}
such that the quantum limit can be cast in the form found in Ref.~\cite{GPZ10}, namely, 
$ k_{\mathrm{B}} T_{\mathrm{nM}} / \hbar \omega \le \mu/\gamma$, but with $ T_{\mathrm{nM}}$
instead of $T$.
It is also possible to define an effective coupling constant 
\begin{equation}
\gamma_{\mathrm{nM}} = \gamma \left( 1 - \frac{1}{1+\Omega^2/\omega^2} \right).
\end{equation}
For $\Omega \sim \omega$, $T_{\mathrm{nM}} \sim \frac{1}{2}T$ and 
$\gamma_{\mathrm{nM}} \sim \frac{1}{2}\gamma$.
This scaling of the temperature or the coupling constant explains the results depicted in 
Figs.~\ref{fig:equal_tempdep} and \ref{fig:equal_gammadep}, respectively.
Alternatively, the non-Markovian scaling factor 
$1 - \frac{1}{1+\Omega^2/\omega^2} $ can be assigned to a third energy scale.
Define 
\begin{equation}
\mu_{\mathrm{nM}} = \mu \left( 1 - \frac{1}{1+\Omega^2/\omega^2} \right)^{-1},
\end{equation}
so that $\mu_{\mathrm{nM}} \ge \mu$ provided by the fact that 
$1 - \frac{1}{1+\Omega^2/\omega^2} \le 1$.
To be more concrete, note that the particular situation analyzed in Ref.~\cite{GPZ10} 
corresponds to the case $J(\omega)/m = \gamma \omega$ and $\mu$ corresponds to 
the imaginary part of the associated Mathieu coefficient.
Assuming that to leading order in the coupling $c_1$ [see Eq.~(\ref{eq-chrntpmpng})] 
the imaginary part of the Mathieu coefficient can still be expressed by 
$\mu \sim c_1/4\omega$, non-Markovian dynamics can be seen as effectively enhancing 
the coupling between oscillators.
This implies that under non-Markovian dynamics, the amplitude of the driving force
needed for the entanglement to survive is clearly smaller than in the Markovian case.

Although the non-Markovianly scaled parameters $T_{\mathrm{nM}} $, $\gamma_{\mathrm{nM}}$ 
and $\mu_{\mathrm{nM}} $ are particular to the spectral density in Eq.~(\ref{eq:Jomega}),
based on the extensively studied features of non-Markovian dynamics 
\cite{IF09,NBT11,PB12c,CP&13,MBF11,CHP12,PT&14,WS&13,SN&11,SSA14}, there is no 
apparent physical reason why not to expect the same scaling scenario in general.

\section{Entanglement Dynamics for Asymmetric Thermal Baths}
The presence of driving forces above already placed the system into a nontrivial out-of-equilibrium
situation.
Another nontrivial out-of-equilibrium situation that this system may encounter is the case of 
environments at different temperatures $T_1 \neq T_2$ or resonators with different couplings 
constants $\gamma_1 \neq \gamma_2$.
Motivated by the possible role that heat currents may play in the generation entanglement, 
these two situations are considered below.
 
\subsection{Thermal baths at different temperature: $T_1 \neq T_2$}
\begin{SCfigure}[][h!]
\centering
\caption{Non-Markovian entanglement dynamics for degenerate oscillators, $m_1 = m_2 = m$, 
$\omega_1= \omega_2 = \omega$, and baths at different temperatures, 
$T_2 \neq T_1$, with $k_{\mathrm{B}} T_1 = 20 \hbar\omega$.
Parameter values are 
$\gamma_1= \gamma_2 =10^{-3}\omega$, $\Omega_1= \Omega_2 = \Omega$, 
$c_1 = 0.2 m \omega^2$, and $\omega_{\mathrm{d}} = 2 \omega$.
}
\includegraphics[width=0.6\columnwidth]{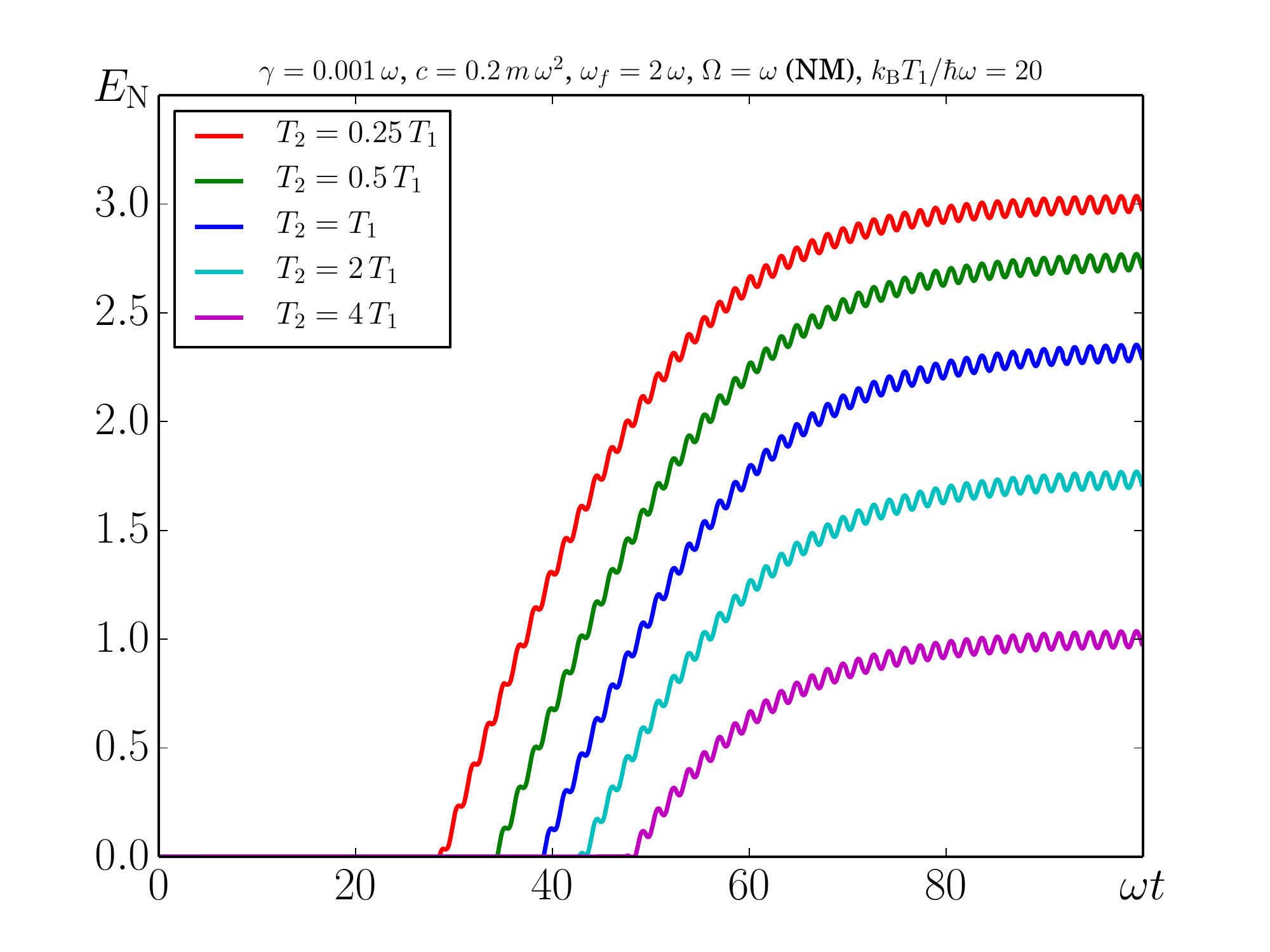}
\label{fig:diff-temp}
\end{SCfigure}

From Fig.~\ref{fig:equal_tempdep} it is clear that the lower the temperature, the 
larger the amount of steady state entanglement that is reached.
However, because heat transfer between thermal baths at different temperature is assisted 
here by the interaction between oscillators, asking for (i) the possible role of heat transfer in 
preserving/destroying quantum correlations between oscillators and for (ii) the classical/quantum 
nature of possible correlations established by heat fluxes at high/low temperature among 
the oscillators, are legitime questions. 
For concreteness of the present work, these concerns are analyzed in a separated contribution
and here interest is restricted to the amount of entanglement that can be reached for different
temperature ratios.
Specifically, Fig.~\ref{fig:diff-temp} depicts the time dynamics of entanglement for a variety 
of temperature ratios $T_2/T_1$ at fixed $T_1$.
The main features in this figure are the strong dependence of the time at which entanglement 
is generated on the temperature ratio and the dependence of entanglement on the absolute value 
of the temperature of the baths and not only on the temperature difference.
This indeed motivates a comprehensive analysis of entanglement dynamics and heat fluxes.

\subsection{Thermal baths with different decay rate: $\gamma_1 \neq \gamma_2$}
\begin{SCfigure}[][h!]
\centering
\caption{Non-Markovian entanglement dynamics for degenerate oscillators, $m_1 = m_2 = m$, 
$\omega_1= \omega_2 = \omega$,  with different coupling constant to the baths,
$\gamma_2 \neq \gamma_1$, with $\gamma_1 = 5\times 10^{-3}\omega$.
Parameter values are $k_{\mathrm{B}}T_1= k_{\mathrm{B}}T_2 = 5 \hbar \omega$, 
$\Omega_1= \Omega_2 = \Omega$, 
$c_1 = 0.2 m \omega^2$, and $\omega_{\mathrm{d}} = 2 \omega$.
}
\includegraphics[width=0.6\columnwidth]{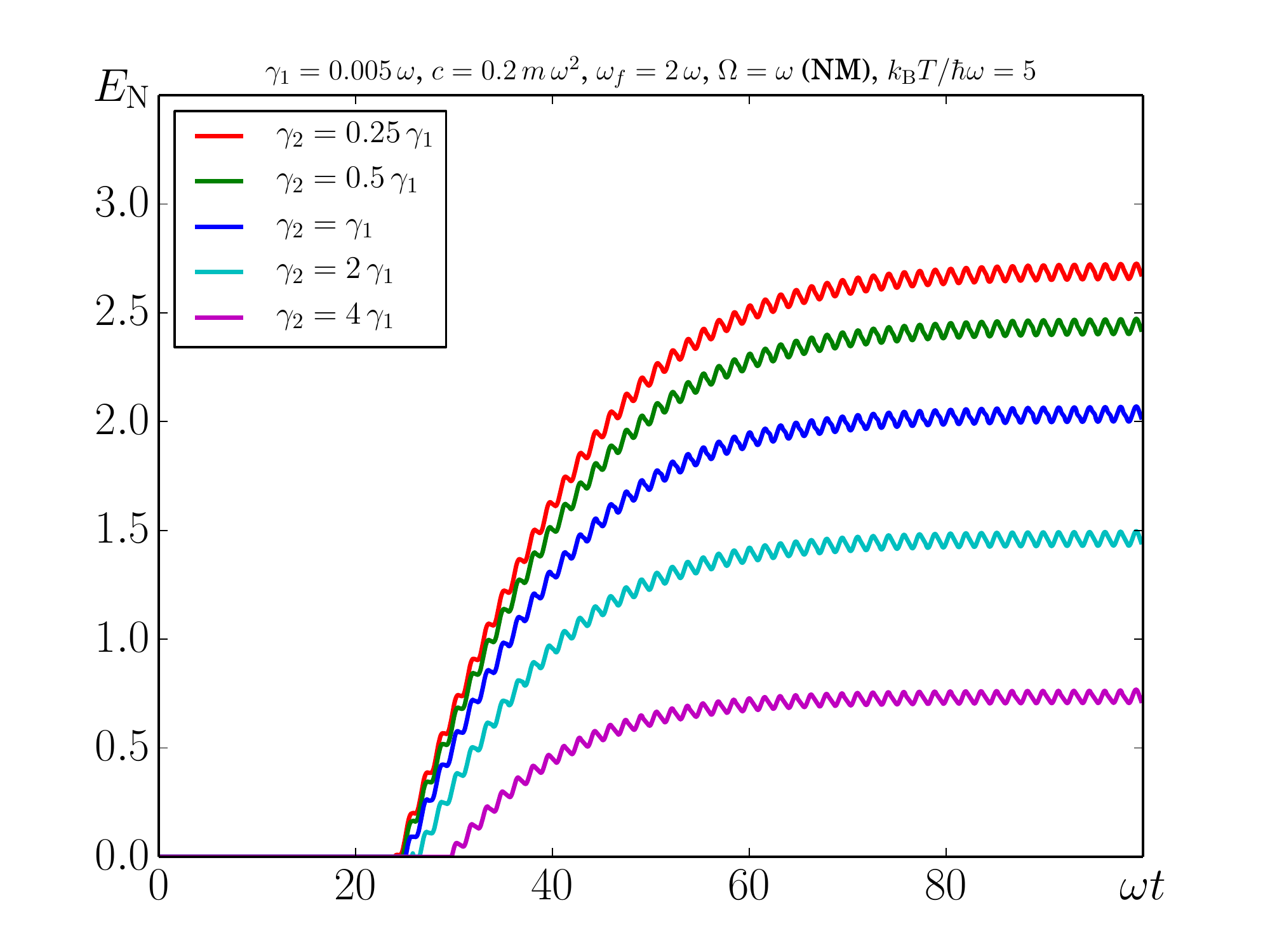}
\label{fig:diff-dissp}
\end{SCfigure}

Because the effective temperature at which each oscillator thermalizes is a function of the
power noise of the bath and therefore of the coupling constant \cite{PT&14}, another interesting
situation from a thermodynamic point of view is the case when the coupling constants are
different.
Figure~\ref{fig:diff-dissp} depicts the time dynamics of entanglement for a variety 
of coupling constant ratios $\gamma_2/\gamma_1$ at fixed $\gamma_1$.
Up to the weak dependence of the time at which entanglement is generated on the coupling ratios,
the behaviour is essentially the same as in Fig.~\ref{fig:diff-temp}.
Below, the case of non-degenerate oscillators is analyzed and the effect of non-Markovian dynamics
in the resonance condition $\omega_{\mathrm{d}} = \omega_1 + \omega_2$ is discussed.

\section{Entanglement Dynamics for Asymmetric Oscillators}
Since the case of degenerate oscillators lacks of experimental relevance \cite{GGZ10} and the 
generation of squeezing in mechanical setups was already achieved for non-degenerate oscillators
\cite{MO&14}, the dynamics of entanglement is considered next for different masses and 
frequencies.
This last situation complements the analysis of the influence of asymmetries started in the 
previous section with the cases $T_1 \neq T_2$ and $\gamma_1 \neq \gamma_2$.

The undriven non-Markovian dynamics for the degenerate and non-degenerate cases were 
previously analyzed in Ref.~\cite{GGZ10}.
In particular, Ref.~\cite{GGZ10} addresses the effect of the resonance condition for degenerate 
oscillators and its relationship with the possibility of preserving  entanglement at asymptotic times.

\subsection{Entanglement dynamics for oscillators of different natural frequency $\omega_1 
\neq \omega_2$}
For small values of $c_1$ and in the rotating wave approximation, the maximum rate of
generation of squeezing is obtained for $\omega_{\mathrm{d}}= \omega_1+\omega_2$. 
For the degenerate case, it reads $\omega_{\mathrm{d}}= 2\omega_1$.
\begin{SCfigure}[10][h!]
\centering
\caption{Non-Markovian entanglement dynamics for equal mass resonators 
$m_1=m_2=m$ but with $\omega_1 \neq \omega_2$ for  
$\Omega_1=\Omega_2=\omega_1$.
Parameter values are $k_{\mathrm{B}}T_1= k_{\mathrm{B}}T_2= 60 \hbar \omega_1$,
$\gamma_1=\gamma_2 = 10^{-3}\omega_1$, $c_1 = 0.2 m \omega_1^2$, 
and $\omega_{\mathrm{d}} = 2 \omega_1$.}
\includegraphics[width=0.6\columnwidth]{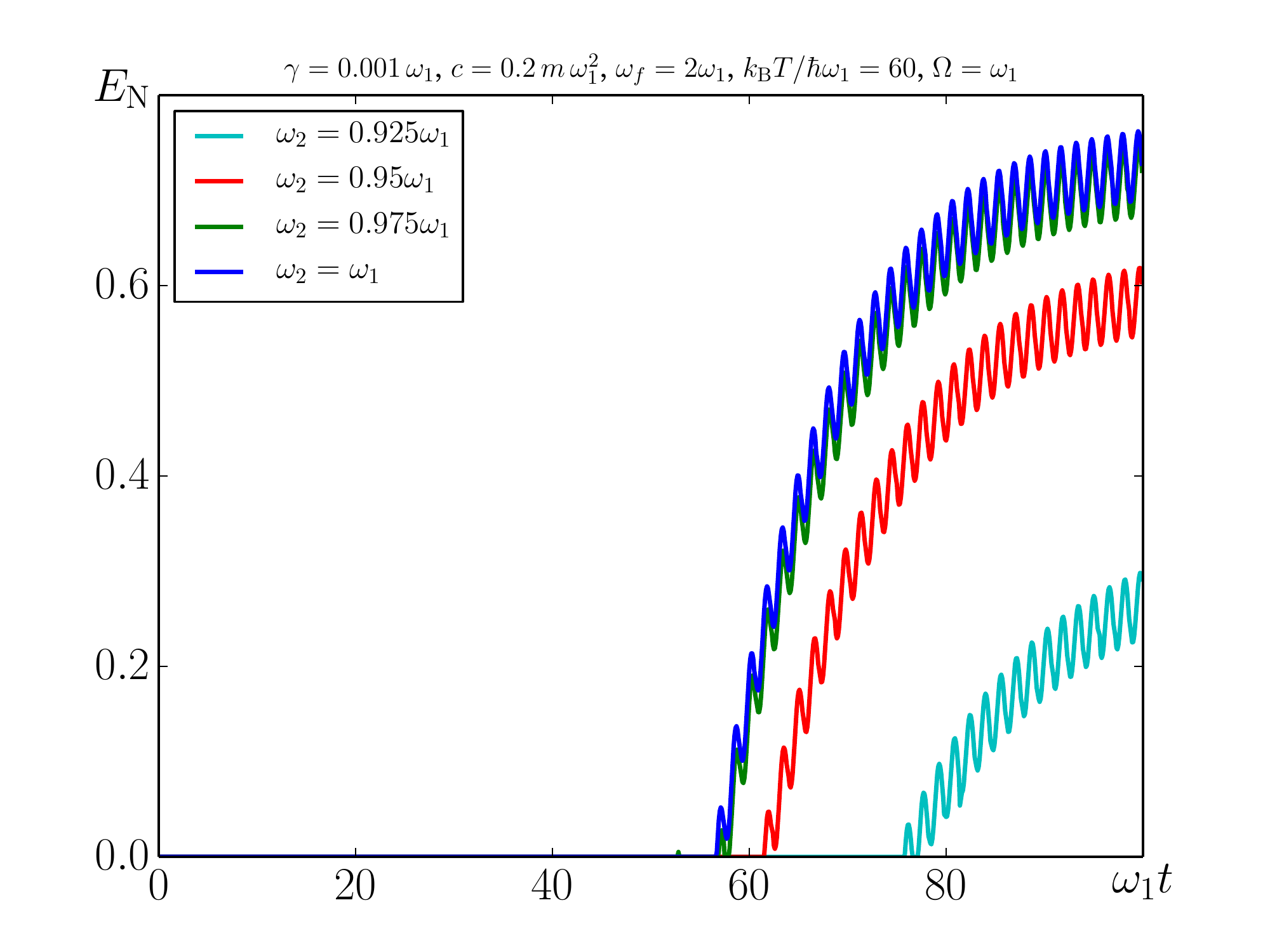}
\label{fig:asymm_omega}
\end{SCfigure}
Below, for this optimal condition, the dynamics of entanglement is analyzed for a variety
of frequency ratios $\omega_2/\omega_1$ for fixed $\omega_1$.
To isolate the effect of non-Markovian dynamics, parameters are chosen so that no entanglement
is found in the Markovian degenerate case with $\omega_{\mathrm{d}}= 2\omega_1$.
Figure~\ref{fig:asymm_omega} not only shows that non-Markovian dynamics support the 
creation and survival of steady entanglement for the degenerate case  with ``resonant driving'', 
but also over a broad range of frequency detuning ($\sim7\%$) and with ``non-resonant'' driving.
This feature adds to the known robustness of the squeezing-entanglement generation against 
small detuning form the resonance frequency. 

In the experimental work on the generation of two-mode squeezing reported in Ref.~\cite{MO&14}, 
$\omega_2 \sim 0.939 \omega_1$ with $\omega_{\mathrm{d}} = \omega_1 + \omega_2$
and $\gamma \sim 10^{-3}$, so that the for non-Markovian scenario and pumping rate in 
Fig.~\ref{fig:asymm_omega}, entanglement can be reached at $\sim 37$~mK.
For quality factors two orders of magnitude larger, entangled modes could be found at $3.7$~K.
Because entanglement is not found for this set of parameters in the Markovian case, note that 
reaching these high temperatures is supported by the non-Markovian character of the dynamics.

\subsection{Entanglement dynamics for oscillators of different mass $m_1 \neq m_2$}
\begin{SCfigure}[10][h!]
\centering
\caption{Non-Markovian entanglement dynamics for equal frequency resonators 
$\omega_1=\omega_2=\omega$ but with $m_1 \neq m_2$ for  $\Omega_1=\Omega_2=\omega$.
Parameter values are $k_{\mathrm{B}}T_1= k_{\mathrm{B}}T_2= 60 \hbar \omega$,
$\gamma_1=\gamma_2 = 10^{-3}\omega$, $c_1 = 0.2 m \omega^2$, 
and $\omega_{\mathrm{d}} = 2 \omega$.}
\includegraphics[width=0.6\columnwidth]{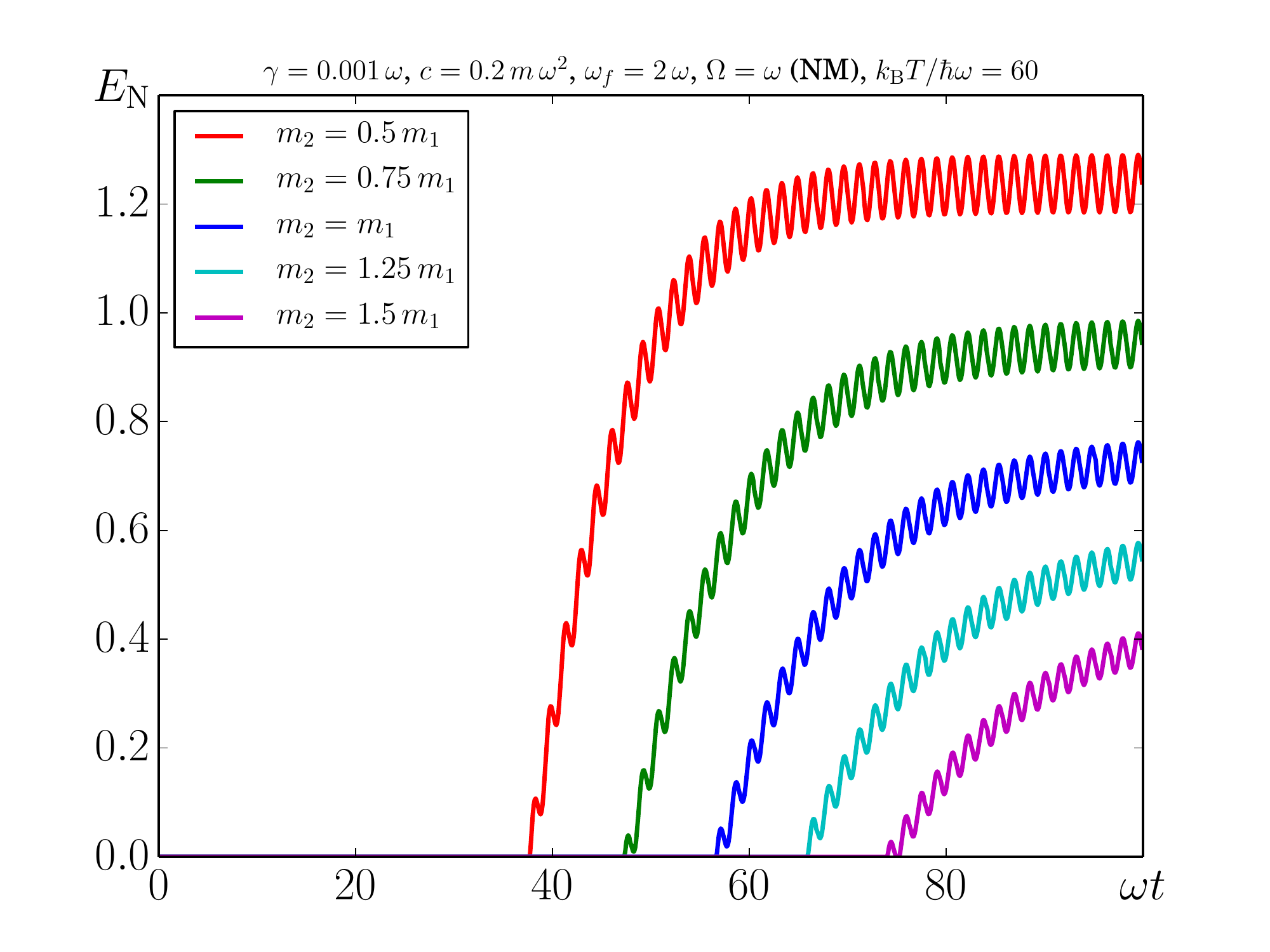}
\label{fig:asymm_mass}
\end{SCfigure}

Figure \ref{fig:asymm_mass} depicts  the logarithmic negativity for a 
variety of mass ratios $m_2/m_1$ for fixed $m_1$.
As it is expected, the smaller the ratio  $m_2/m_1$  is, the more effective the driving field is
in generating entanglement out of the modulation of the coupling strength.
Note that the smaller the mass ratio is, not only the larger the value of $E_\mathrm{N}$ is, but also the 
shorter the time at which entanglement is generated.
This is consistent with the intuitive ideas about effectiveness of the driving field in the presence 
of lighter masses.
%

\section{Discussion}
For highly symmetric cases, there is evidence that non-Markovian dynamics may allow for the 
survival of entanglement at temperatures higher  than the corresponding Markovian case provided 
by the interaction with a common bath \cite{GGZ10,RR13}.
For purely dephasing baths, non-Markovian dynamics may allow for larger values of entanglement 
\cite{HRP12}.
Based on analytic exact results for the non-Markovian and out-of-equilibrium dynamics of two 
non-degenerate parametrically coupled harmonic oscillators, it is shown here that in the presence 
of time-dependent external fields, non-Markovian dynamics support the generation of out-of-equilibrium 
steady state entanglement at higher temperatures, larger coupling-to-the-environment constants 
and lower pumping rates than in the Markovian case.
This delicate interplay between driving fields and non-Markovian dynamics sets a new quantum
limit which incorporates the main time and energy scales of the physical systems at hand.

\section*{Acknowlegedments}

\noindent
Fruitful and helpful discussions with David Zueco and Fernando Galve are acknowledged
with pleasure.
This work was supported by the \textit{Comit\'e para el Desarrollo de la Investigaci\'on}
--CODI-- of Universidad de Antioquia, Colombia under contract number E01651 
and the Estrategia de Sostenibilidad 2013-2014 and by the \textit{Departamento Administrativo 
de Ciencia, Tecnolog\'ia e Innovaci\'on} --COLCIENCIAS-- of Colombia under the grant number 
111556934912.

\section*{References}
\bibliographystyle{unsrt}
\bibliography{nmql}

\newpage

\appendix
\section{Influence Functional and Propagating Function of the System}
\label{app:InfncFnctnl}

The starting point in the influence-functional theory is considering the density operator 
of the global system at time $t$, that in terms of the initial density operator $\hat{\rho}(0)$ is 
given by
\begin{equation}
\label{eq-6}
\hat{\rho}(t)=\exp\left(-\frac{\mathrm{i}}{\hbar}\hat{H}\,t\right)\hat{\rho}(0)\exp\left(\frac{\mathrm{i}}{\hbar}\hat{H}\,\right),
\end{equation}
where $\hat{H} = \hat{H}_{\mathrm{S}} + \hat{H}_{\mathrm{B}} + \hat{H}_{\mathrm{I}}$ is the 
Hamiltonian of the global system with $\hat{H}_{\mathrm{S}}$, $\hat{H}_{\mathrm{B}}$ and 
$\hat{H}_{\mathrm{SB}}$ given by Eqs.~(\ref{eq:SystmHmltnn}) and (\ref{eq:BathSystmHmltnn}).
Physically, expression~(\ref{eq-6}) implies that the global system is considered as a closed 
one, and therefore, it is possible to evolve it in time by means of an unitary operator. 
In the coordinate representation, the density operator $\hat{\rho}(t)$ reads
\begin{equation}
\label{eq-7}
\begin{split}
&\braket{q_{1+}'',q_{2+}'',\vtr{q}_{1+}'',\vtr{q}_{2+}''}{\hat{\rho}(t)}{q_{1-}'',q_{2-}'',\vtr{q}_{1-}'',\vtr{q}_{2-}''} = 
 \\
&\phantom{f} \int_{-\infty}^\infty\df{q_{1+}'}\df{q_{2+}'}\df{\vtr{q}_{1+}'}\df{\vtr{q}_{2+}'}
\df{q_{1-}'}\df{q_{2-}'}\df{\vtr{q}_{1-}'}\df{\vtr{q}_{2-}'}
\braket{q_{1+}'',q_{2+}'',\vtr{q}_{1+}'',\vtr{q}_{2+}''}{\mathrm{e}^{-\frac{\mathrm{i}}{\hbar}\hat{H}\,t}}
{q_{1+}',q_{2+}',\vtr{q}_{1+}',\vtr{q}_{2+}'} 
 \\
&\phantom{f} \times 
\braket{q_{1+}',q_{2+}',\vtr{q}_{1+}',\vtr{q}_{2+}'}{\hat{\rho}(0)}{q_{1-}',q_{2-}',\vtr{q}_{1-}',\vtr{q}_{2-}'}
\braket{q_{1-}',q_{2-}',\vtr{q}_{1-}',\vtr{q}_{2-}'}{\mathrm{e}^{\frac{\mathrm{i}}{\hbar}\hat{H}\,t}}
{q_{1-}'',q_{2-}'',\vtr{q}_{1-}'',\vtr{q}_{2-}''},
\end{split}
\end{equation}
where $q_{\alpha\pm}$ stands for the coordinates of the oscillators in the system of interest, and  
$\vtr{q}_{\alpha\pm}=(q_{\alpha,1\pm},q_{\alpha,2\pm},\dots,q_{\alpha,N\pm})$ the coordinates of 
the thermal baths. 
The matrix elements for the temporal evolution operators are known as the \textit{propagating 
kernel}\cite{Wei08,FV63,GSI88,Ing02} and is given by
\begin{equation}
\label{eq-8}
\begin{split}
&\braket{q_{1+}'',q_{2+}'',\vtr{q}_{1+}'',\vtr{q}_{2+}''}{\mathrm{e}^{-\frac{\mathrm{i}}{\hbar}\hat{H}\,t}}
{q_{1+}',q_{2+}',\vtr{q}_{1+}',\vtr{q}_{2+}'}= 
\\
&\phantom{f(a+b)} =K(q_{1+}'',q_{2+}'',\vtr{q}_{1+}'',\vtr{q}_{2+}'',t;q_{1+}',q_{2+}',\vtr{q}_{1+}',\vtr{q}_{2+}',0) 
\\
&\phantom{f(a+b)}=\int\dfp{q_{1+}}\dfp{q_{2+}}\dfp{\vtr{q}_{1+}}\dfp{\vtr{q}_{2+}}
\exp\left(\frac{\mathrm{i}}{\hbar}S[q_{1+},q_{2+},\vtr{q}_{1+},\vtr{q}_{2+}]\right).
\end{split}
\end{equation}
This object evolves the overall system ``forward in time'', that is why the ``$+$'' subscript is used. 
An analogous expression stands for the matrix elements 
$\braket{q_{1-}',q_{2-}'',\vtr{q}_{1-}',\vtr{q}_{2-}'}{\mathrm{e}^{\frac{\mathrm{i}}{\hbar}\hat{H}\,t}}
{q_{1-}'',q_{2-}'',\vtr{q}_{1-}'',\vtr{q}_{2-}''}$
that evolves the overall system ``backward in time'' and this is reflected by the 
``$-$'' subscript in the coordinates.
The path integrals in the propagating kernels must be evaluated over the paths $q_{\alpha\pm}$ 
and $\vtr{q}_{\alpha\pm}$ that satisfy the following boundary conditions
\bea
\label{eq-10}
q_{\alpha\pm}=\begin{cases} q_{\alpha\pm}', & s=0, \\ q_{\alpha\pm}'', & s=t; \end{cases} \quad
\vtr{q}_{\alpha\pm}=\begin{cases} \vtr{q}_{\alpha\pm}', & s=0, \\ \vtr{q}_{\alpha\pm}'', & s=t. 
\end{cases}
\eea
In equation~(\ref{eq-8}), $S$ stands for the action for the global system, 
defined as usual:
\begin{align}\label{eq-11}
S &= \int_0^t\df{s}L_\mathrm{S}(\dot{q}_1(s),\dot{q}_2(s),q_1(s),q_2(s))
+ 
\int_0^t\df{s}L_\mathrm{B}(\dot{\vtr{q}}_1(s),\dot{\vtr{q}}_2(s),\vtr{q}_1(s),
\vtr{q}_2(s))
\notag \\
&+ \int_0^t\df{s}L_\mathrm{I}(q_1(s),q_2(s),\vtr{q}_1(s),\vtr{q}_2(s))
= S_\mathrm{S} + S_\mathrm{B} + S_\mathrm{I},
\end{align}
where $L_\mathrm{S}$, $L_\mathrm{B}$ and $L_\mathrm{I}$ stand for the Lagrangian of the 
system of interest, the thermal baths and the interaction between these subsystems, respectively. 
The Lagrangian have the following form
\begin{align}
L_\mathrm{S} &= 
\sum_{\alpha=1}^2\left(\frac{1}{2}m_\alpha\dot{q}_\alpha^2-\frac{1}{2}m_\alpha\omega_\alpha^2q_\alpha^2\right), 
\label{eq-12} \\
L_\mathrm{B} &= \sum_{\alpha=1}^2\sum_{k=1}^N\left(\frac{1}{2}m_{\alpha,k}\dot{q}_{\alpha,k}^2-
\frac{1}{2}m_{\alpha,k}\omega_{\alpha,k}^2q_{\alpha,k}^2\right), \label{eq-13} \\
L_\mathrm{I} &= \sum_{\alpha=1}^2\sum_{k=1}^N\left(q_{\alpha}\,c_{\alpha,k}\,q_{\alpha,k}-
q_\alpha^2\frac{c_{\alpha,k}^2}{2m_{\alpha,k}\omega_{\alpha,k}^2}\right). \label{eq-14}
\end{align}
By replacing the expression (\ref{eq-8}) into (\ref{eq-7}), the matrix elements of the density 
operator of the global system read
\begin{align}
\label{eq-15}
&\braket{q_{1+}'',q_{2+}'',\vtr{q}_{1+}'',\vtr{q}_{2+}''}{\hat{\rho}(t)}{q_{1-}'',q_{2-}'',\vtr{q}_{1-}'',\vtr{q}_{2-}''} = 
\notag \\
&\phantom{f} \int_{-\infty}^\infty\df{q_{1+}'}\df{q_{2+}'}\df{\vtr{q}_{1+}'}\df{\vtr{q}_{2+}'}
\df{q_{1-}'}\df{q_{2-}'}\df{\vtr{q}_{1-}'}\df{\vtr{q}_{2-}'}
K(q_{1+}'',q_{2+}'',\vtr{q}_{1+}'',\vtr{q}_{2+}'',t;q_{1+}',q_{2+}',\vtr{q}_{1+}',\vtr{q}_{2+}',0) 
\notag \\
&\phantom{f} \times K^*(q_{1-}'',q_{2-}'',\vtr{q}_{1-}'',\vtr{q}_{2-}'',t;q_{1-}',q_{2-}',\vtr{q}_{1-}',\vtr{q}_{2-}',0)
\braket{q_{1+}',q_{2+}',\vtr{q}_{1+}',\vtr{q}_{2+}'}{\hat{\rho}(0)}{q_{1-}',q_{2-}',\vtr{q}_{1-}',\vtr{q}_{2-}'}.
\end{align}
This expression describes the dynamics of the global system. 
However, all this information is not necessary. 
The only information that is relevant for the present case is that of the system of interest 
under the influence on the environment.
Therefore, the relevant object here is the reduced density matrix, which is obtained 
by tracing out over the degrees of freedom of the thermal baths in Eq.~(\ref{eq-15}),
i.e.,
\begin{align}
\label{eq-16}
\rho_{\mathrm{S}}(q_{1+}'',q_{2+}'',q_{1-}'',q_{2-}'',t) &= 
\int_{-\infty}^\infty\df{\vtr{q}_{1+}''}\df{\vtr{q}_{2+}''}
\braket{q_{1+}'',q_{2+}'',\vtr{q}_{1+}'',\vtr{q}_{2+}''}{\hat{\rho}(t)}{q_{1-}'',q_{2-}'',\vtr{q}_{1+}'',\vtr{q}_{2+}''} 
\notag \\
&= \int_{-\infty}^\infty\df{\vtr{q}_{1+}''}\df{\vtr{q}_{2+}''}\df{q_{1+}'}\df{q_{2+}'}\df{\vtr{q}_{1+}'}\df{\vtr{q}_{2+}'}
\df{q_{1-}'}\df{q_{2-}'}\df{\vtr{q}_{1-}'}\df{\vtr{q}_{2-}'} 
\notag \\
&\times K(q_{1+}'',q_{2+}'',\vtr{q}_{1+}'',\vtr{q}_{2+}'',t;q_{1+}',q_{2+}',\vtr{q}_{1+}',\vtr{q}_{2+}',0) 
\notag \\
&\times K^*(q_{1-}'',q_{2-}'',\vtr{q}_{1+}'',\vtr{q}_{2+}'',t;q_{1-}',q_{2-}',\vtr{q}_{1-}',\vtr{q}_{2-}',0) 
\notag \\
&\times \braket{q_{1+}',q_{2+}',\vtr{q}_{1+}',\vtr{q}_{2+}'}{\hat{\rho}(0)}{q_{1-}',q_{2-}',\vtr{q}_{1-}',\vtr{q}_{2-}'}.
\end{align}

At this point, it is assumed that the initial density operator of the global system factorizes,
$
\hat{\rho}(0)=\hat{\rho}_\mathrm{S}(0)\otimes\hat{\rho}_{\mathrm{B}_1}(0)\otimes\hat{\rho}_{\mathrm{B}_2}(0).
$
In the position representation, $\hat{\rho}(0)$ reads
\begin{align}\label{eq-18}
&\braket{q_{1+}',q_{2+}',\vtr{q}_{1+}',\vtr{q}_{2+}'}{\hat{\rho}(0)}{q_{1-}',q_{2-}',\vtr{q}_{1-}',\vtr{q}_{2-}'} 
\notag \\
&\phantom{f(a+b)} =
\braket{q_{1+}',q_{2+}'}{\hat{\rho}_\mathrm{S}(0)}{q_{1-}',q_{2-}'}
\braket{\vtr{q}_{1+}'}{\hat{\rho}_{\mathrm{B}_1}(0)}{\vtr{q}_{1-}'}
\braket{\vtr{q}_{2+}'}{\hat{\rho}_{\mathrm{B}_2}(0)}{\vtr{q}_{2-}'} 
\notag \\
&\phantom{f(a+b)} =
\rho_\mathrm{S}(q_{1+}',q_{2+}',q_{1-}',q_{2-}',0)
\rho_{\mathrm{B}_1}(\vtr{q}_{1+}',\vtr{q}_{1-}',0)
\rho_{\mathrm{B}_2}(\vtr{q}_{2+}',\vtr{q}_{2-}',0).
\end{align}
By replacing this expression in Eq.~(\ref{eq-16}), the reduced density matrix is found 
to read
\begin{align}
\label{eq-19}
&\rho_{\mathrm{S}}(q_{1+}'',q_{2+}'',q_{1-}'',q_{2-}'',t) = \notag \\
&\phantom{f(a+b)}\int_{-\infty}^\infty\df{q_{1+}'}\df{q_{2+}'}\df{q_{1-}'}\df{q_{2-}'}
J(q_{1+}'',q_{2+}'',q_{1-}'',q_{2-}'',t;q_{1+}',q_{2+}',q_{1-}',q_{2-}',0) \notag \\
&\phantom{f(a+b)} \times \rho_\mathrm{S}(q_{1+}',q_{2+}',q_{1-}',q_{2-}',0),
\end{align}
where
\begin{align}
\label{eq-20}
&J(q_{1+}'',q_{2+}'',q_{1-}'',q_{2-}'',t;q_{1+}',q_{2+}',q_{1-}',q_{2-}',0) = 
\notag \\
&\phantom{f(a+b)}\int\dfp{q_{1+}}\dfp{q_{2+}}\dfp{q_{1-}}\dfp{q_{2-}}
\exp\biggl\{\frac{\mathrm{i}}{\hbar}\left(S_\mathrm{S}[q_{1+},q_{2+}]-S_\mathrm{S}[q_{1-},q_{2-}]\right)\biggr\} 
\notag \\
&\phantom{f(a+b)} \times \mc{F}[q_{1+},q_{2+},q_{1-},q_{2-}]
\end{align}
denotes the \textit{propagating function} of the reduced density matrix and the object 
$\mc{F}[q_{1+},q_{2+},q_{1-},q_{2-}]$ is the \textit{influence functional}\cite{FV63,FH05}
given by
\begin{align}\label{eq-21}
\mc{F}[q_{1+},q_{2+},q_{1-},q_{2-}] &= 
\int_{-\infty}^\infty\df{\vtr{q}_{1+}''}\df{\vtr{q}_{2+}''}\df{\vtr{q}_{1+}'}\df{\vtr{q}_{2+}'}
\df{\vtr{q}_{1-}'}\df{\vtr{q}_{2-}'}\rho_{\mathrm{B}_1}(\vtr{q}_{1+}',\vtr{q}_{1-}',0)
\rho_{\mathrm{B}_2}(\vtr{q}_{2+}',\vtr{q}_{2-}',0) 
\notag \\
&\times \int\dfp{\vtr{q}_{1+}}\dfp{\vtr{q}_{2+}}\dfp{\vtr{q}_{1-}}\dfp{\vtr{q}_{2-}}
\exp\biggl\{\frac{\mathrm{i}}{\hbar}\left(S_\mathrm{B}[\vtr{q}_{1+},\vtr{q}_{2+}]-
S_\mathrm{B}[\vtr{q}_{1-},\vtr{q}_{2-}]\right.\biggr. \notag \\
&+ \biggl.\left. S_\mathrm{I}[q_{1+},q_{2+},\vtr{q}_{1+},\vtr{q}_{2+}]-
S_\mathrm{I}[q_{1-},q_{2-},\vtr{q}_{1-},\vtr{q}_{2-}]\right)\biggr\}.
\end{align}

\subsection{Derivation of the influence functional}
To derive the influence functional, it is useful to use one of the properties of the influence 
functional, specfically,
\begin{quote}
\textit{If a system is interacting simultaneously with two uncoupled and independents environments 
$A$ and $B$, with no initial correlations between them, then}\cite{FV63}
\begin{equation}\label{ap-1}
F=F_A\cdot F_B.
\end{equation}
\end{quote}
Using the above property, it is possible to express the influence functional in (\ref{eq-21}) as 
the product of two influence functionals, one for each  oscillator in the system of interest, 
namely,
\begin{equation}
\label{ap-2}
\mc{F}[q_{1+},q_{2+},q_{1-},q_{2-}]=\mc{F}[q_{1+},q_{1-}]\mc{F}[q_{2+},q_{2-}],
\end{equation}
where
\begin{align}
\label{ap-3}
\mc{F}[q_{\alpha+},q_{\alpha-}] &= \int_{-\infty}^\infty\df{\vtr{q}_{\alpha+}''}
\df{\vtr{q}_{\alpha+}'}\df{\vtr{q}_{\alpha-}'}
 \rho_{\mathrm{B}_\alpha}(\vtr{q}_{\alpha+}',\vtr{q}_{\alpha-}',0) 
 \notag \\
 &\times \int\dfp{\vtr{q}_{\alpha+}}\dfp{\vtr{q}_{\alpha-}}
 \exp\biggl\{\frac{\mathrm{i}}{\hbar}\biggl(S_{\mathrm{B}_\alpha}[\vtr{q}_{\alpha+}]+
 S_{\mathrm{I}_\alpha}[\vtr{q}_{\alpha+},\vtr{q}_{\alpha+}]\biggr.\biggr. 
 \notag \\
 &- S_{\mathrm{B}_\alpha}[\vtr{q}_{\alpha-}]
 -S_{\mathrm{I}_\alpha}[q_{\alpha-},\vtr{q}_{\alpha-}]\biggl.\biggl.\biggr)\biggr\}, 
 \quad \alpha=1,2.
\end{align}
Having in mind that the oscillators encompassing each thermal bath are independent 
among them, the property of the influence functional stated above can be used once 
more so that each influence functional in Eq.~(\ref{ap-2}) is given by
\begin{equation}
\label{ap-4}
\mc{F}[q_{\alpha+},q_{\alpha-}]=\prod_{k=1}^N\mc{F}_k[q_{\alpha+},q_{\alpha-}],
\end{equation}
where each  $\mc{F}_k[q_{\alpha+},q_{\alpha-}]$ describes the influence of each oscillator 
in one thermal bath on the oscillator in the system of interest.
Each of these influence functionals reads
\begin{align}
\label{ap-5}
\mc{F}_k[q_{\alpha +},q_{\alpha -}] &= \int_{-\infty}^\infty \df{q_{\alpha,k +}''}\df{q_{\alpha,k
+}'}\df{q_{\alpha,k -}'}\rho_{\mathrm{B}_\alpha}^{(k)}(q_{\alpha,k +}',q_{\alpha,k -}',0) 
\notag \\
&\times \int\dfp{q_{\alpha,k +}}\exp\left\{\frac{\mathrm{i}}{\hbar}
\left(S_{\mathrm{B}_\alpha}^{(k)}[q_{\alpha,k
+}] + S_{\mathrm{I}_\alpha}^{(k)}[q_{\alpha +},q_{\alpha,k +}]\right)\right\} 
\notag \\
&\times \int\dfp{q_{\alpha,k -}}\exp\left\{-\frac{\mathrm{i}}{\hbar}
\left(S_{\mathrm{B}_\alpha}^{(k)}[q_{\alpha,k
-}] + S_{\mathrm{I}_\alpha}^{(k)}[q_{\alpha -},q_{\alpha,k -}]\right)\right\}.
\end{align}
$\rho_{\mathrm{B}_\alpha}^{(k)}(q_{\alpha,k +}',q_{\alpha,k -}',0)$ 
denotes the initial density matrix for the $k^{\mathrm{th}}$ oscillator in the thermal bath. 
For the case of a thermal bath at thermal equilibrium at a temperature $T_\alpha$, it 
reads (see, e.g., Ref.~\cite{Ing02})
\begin{align}
\label{ap-6}
&\rho_{\mathrm{B}_\alpha}^{(k)}(q_{\alpha,k +}',q_{\alpha,k -}',0) 
\notag \\
&\phantom{f} = 2\sinh\left(\frac{\hbar \beta_\alpha
\omega_{\alpha,k}}{2}\right)
\sqrt{\frac{m_{\alpha,k}\omega_{\alpha,k}}{2\pi\hbar\sinh(\hbar\beta_\alpha\omega_{\alpha,k})}} 
\notag \\
&\phantom{f} \times 
\exp\left\{-\frac{m_{\alpha,k}\omega_{\alpha,k}}{2\hbar\sinh(\hbar\beta_\alpha\omega_{\alpha,k})}
\left[\left(q_{\alpha,k +}'^2+q_{\alpha,k -}'^2\right)\cosh(\hbar\beta_\alpha\omega_{\alpha,k})
-2q_{\alpha,k +}'q_{\alpha,k -}'\right]\right\}.
\end{align}

The next step is to evaluate the path integrals in Eq.~(\ref{ap-5}). 
In doing so, note that the global system under study is linear, 
and therefore, it is only necessary to evaluate the actions $S_{\mathrm{B}_\alpha}^{(k)}$ and 
$S_{\mathrm{I}_\alpha}^{(k)}$ along the classical paths $q_{\alpha,k}(s)$ of each oscillator 
in the thermal baths. 
These classical paths are solution to the  equation of motion
\begin{equation}
\label{ap-7}
\ddot{q}_{\alpha,k}(s)+
\omega_{\alpha,k}^2q_{\alpha,k}(s)=q_\alpha(s)\frac{c_{\alpha,k}}{m_{\alpha,k}},
\end{equation}
that can be obtained from the Lagrangians in (\ref{eq-13}) and (\ref{eq-14}). 
The trick for solving this differential equation consists in treating the system coordinate 
$q_\alpha$ as if it were a given function of time. 
So that, a differential equation for a driven harmonic oscillator is obtained. 
Therefore, the path integrals in (\ref{ap-5}) correspond to the kernel for a driven harmonic 
oscillator (see, e.g., Ref.~\cite{Ing02}), except for the term 
\begin{equation*}
q_\alpha^2\frac{c_{\alpha,k}^2}{2m_{\alpha,k}\omega_{\alpha,k}^2}.
\end{equation*}
This contribution can be taken out of the path integral because it does not contain any term 
that depends on the classical paths of the bath oscillators.
Having in mind the boundary conditions in Eq.~(\ref{eq-10}), taking 
$\vtr{q}_{\alpha -}(t)=\vtr{q}_{\alpha +}''$ for the tracing operation and using the 
propagation kernel for a driven harmonic oscillator, the path integrals in (\ref{ap-5}) 
are readily given by
\begin{align}
\label{ap-8}
&\int\dfp{q_{\alpha,k \pm}}
\exp\left\{\pm\frac{\mathrm{i}}{\hbar}\left(S_{\mathrm{B}_\alpha}^{(k)}[q_{\alpha,k\pm}]+
S_{\mathrm{I}_\alpha}^{(k)}[q_{\alpha \pm},q_{\alpha,k \pm}]\right)\right\}  
\notag \\
&\phantom{f} =
\sqrt{\frac{m_{\alpha,k }\omega_{\alpha,k }}{2\pi(\pm\mathrm{i})\hbar\sin(\omega_{\alpha,k }t)}}
\exp\biggl(\pm\frac{\mathrm{i}}{\hbar}\biggl\{\frac{m_{\alpha,k }\omega_{\alpha,k }}{2\sin(\omega_{\alpha,k}t)}
\biggl[
\left(q_{\alpha,k +}''^2+q_{\alpha,k \pm}'^2\right)\cos(\omega_{\alpha,k }t)\biggr.\biggr.\biggr. 
\notag \\
&\phantom{f} \biggl.-2q_{\alpha,k +}''q_{\alpha,k\pm}'\biggr]+\frac{c_{\alpha,k}q_{\alpha,k
+}''}{\sin(\omega_{\alpha,k}t)}\int_0^t\df{s}\sin(\omega_{\alpha,k}s)q_{\alpha\pm}(s) 
\notag \\
&\phantom{f} + \frac{c_{\alpha,k}q_{\alpha,k\pm}'}{\sin(\omega_{\alpha,k}t)}
\int_0^t\df{s}\sin[\omega_{\alpha,k}(t-s)]q_{\alpha\pm}(s)-\frac{c_{\alpha,k}^2}{2m_{\alpha,k}\omega_{\alpha,k}^2}
\int_0^t\df{s}q_{\alpha\pm}^2(s) 
\notag \\
&\phantom{f} \biggl.\biggl.-\frac{c_{\alpha,k}^2}{m_{\alpha,k}\omega_{\alpha,k}\sin(\omega_{\alpha,k}t)}
\int_0^t\df{s}\int_0^s\df{u}\sin(\omega_{\alpha,k}u)\sin[\omega_{\alpha,k}(t-s)]q_{\alpha\pm}(s)q_{\alpha\pm}(u)
\biggr\}\biggr).
\end{align}
Using this expression and Eq.~(\ref{ap-6}), the influence functional in Eq.~(\ref{ap-5}) takes the form
\begin{align}
\label{ap-9}
&\mc{F}[q_{\alpha,+},q_{\alpha,-}] 
\notag \\
&\phantom{f} =2\sinh\left(\frac{\hbar\beta_\alpha\omega_{\alpha,k}}{2}\right)
\sqrt{\frac{m_{\alpha,k}\omega_{\alpha,k}}{2\pi\hbar\sinh(\hbar\beta_\alpha\omega_{\alpha,k})}}
\sqrt{\frac{m_{\alpha,k}\omega_{\alpha,k}}{2\pi \mathrm{i}\hbar\sin(\omega_{\alpha,k}t)}}
\sqrt{\frac{m_{\alpha,k}\omega_{\alpha,k}\mathrm{i}}{2\pi \hbar\sin(\omega_{\alpha,k}t)}} 
\notag \\
&\phantom{f}\times \int_{-\infty}^\infty\df{q_{\alpha,k+}''}\df{q_{\alpha,k+}'}\df{q_{\alpha,k-}'}
\exp\biggl\{-\frac{m_{\alpha,k}\omega_{\alpha,k}}{2\hbar\sinh(\hbar\beta_\alpha\omega_{\alpha,k})}
\biggl[\left(q_{\alpha,k+}'^2+q_{\alpha,k-}'^2\right)\biggr.\biggr. 
\notag \\
&\phantom{f} \times\biggl.\cosh(\hbar\beta_\alpha\omega_{\alpha,k})-2q_{\alpha,k+}'q_{\alpha,k-}'\biggr]
+\frac{\mathrm{i}}{\hbar}\biggl\{\frac{m_{\alpha,k}\omega_{\alpha,k}}{2\sin(\omega_{\alpha,k}t)}
\biggl[\left(q_{\alpha,k+}'^2-q_{\alpha,k-}'^2\right)\cos(\omega_{\alpha,k}t)\biggr.\biggr. 
\notag \\
&\phantom{f} -\biggl.2q_{\alpha,k+}''(q_{\alpha,k+}'-q_{\alpha,k-}')\biggr]+
\frac{c_{\alpha,k}q_{\alpha,k+'}}{\sin(\omega_{\alpha,
k}t)}\int_0^t\df{s}\sin[\omega_{\alpha,k}(t-s)]q_{\alpha +}(s) 
\notag \\
&\phantom{f} -\frac{c_{\alpha,k}q_{\alpha,k-}'}{\sin(\omega_{\alpha,k}t)}
\int_0^t\df{s}\sin[\omega_{\alpha,k}(t-s)]q_{\alpha-}(s) -
\frac{c_{\alpha,k}^2}{2m_{\alpha,k}\omega_{\alpha,k}^2}
\int_0^t\df{s}\left[q_{\alpha +}^2(s)-q_{\alpha-}^2(s)\right] 
\notag \\
&\phantom{f} + \frac{c_{\alpha,k}q_{\alpha,k+}''}{\sin(\omega_{\alpha,k}t)}
\int_0^t\df{s}\sin(\omega_{\alpha,k}s)
\left[q_{\alpha +}(s)-q_{\alpha -}(s)\right] -\frac{c_{\alpha,k}^2}{m_{\alpha,k}\omega_{\alpha,k}\sin(\omega_{\alpha,k}t)} 
\notag \\
&\phantom{f} \times \biggl.
\int_0^t\df{s}\int_0^s\df{u}\sin(\omega_{\alpha,k}u)\sin[\omega_{\alpha,k}(t-s)]
\left[q_{\alpha +}(s)q_{\alpha +}(u)-q_{\alpha -}(s)q_{\alpha -}(u)\right]\biggr\}.
\end{align}
After performing the integrations over $q_{\alpha,k+}''$, $q_{\alpha,k+}'$, $q_{\alpha,k-}'$, the
influence functional reads
\begin{align}\
\label{ap-10}
&\mc{F}_k[q_{\alpha +},q_{\alpha -}] 
\notag \\
&\phantom{f} = \exp\biggl(-\frac{\mathrm{i}}{\hbar}\biggl\{(q_{\alpha +}'+q_{\alpha -}')
\int_0^t\df{s}\frac{c_{\alpha ,k}^2}{2m_{\alpha,k}\omega_{\alpha ,k}^2}\cos(\omega_{\alpha ,k}s)
\left[q_{\alpha +}(s)-q_{\alpha -}(s)\right]\biggr.\biggr. 
\notag \\
&\phantom{f} + \biggl.\biggl.
\int_0^t\df{s}\int_0^s\df{u}\frac{c_{\alpha ,k}^2}{2m_{\alpha,k}\omega_{\alpha
,k}^2}\cos[\omega_{\alpha ,k}(s-u)]
\left[\dot{q}_{\alpha +}(u)+\dot{q}_{\alpha-}(u)\right]\left[q_{\alpha +}(s)-q_{\alpha -}(s)\right]\biggr\}\biggr)
\notag \\
&\phantom{f} \times \exp\biggl\{-\frac{1}{\hbar}\frac{c_{\alpha,k}^2}{2m_{\alpha,k}\omega_{\alpha,k}}
\int_0^t\df{s}\int_0^s\df{u}
\coth\left(\frac{\hbar\beta_\alpha\omega_{\alpha,k}}{2}\right)\cos[\omega_{\alpha,k}(u-s)]\biggr.
\notag \\
&\phantom{f} \times \left[q_{\alpha +}(s)-q_{\alpha -}(s)\right]\left[q_{\alpha +}(u)-q_{\alpha -}(u)\right]\biggl.\biggr\}.
\end{align}
The above expression describes the influence of the $k-{\mathrm{th}}$ 
oscillator in the $\alpha-{\mathrm{th}}$ thermal bath on the $\alpha-{\mathrm{th}}$ 
oscillator in the system of interest. 
By replacing this expression in (\ref{ap-4}), the complete expression for the influence functional 
is obtained for one thermal bath. 
Specifically, 
\begin{align}
\label{ap-11}
&\mc{F}[q_{\alpha +},q_{\alpha -}]
\notag \\
&\phantom{f} =\exp\biggl(-\frac{\mathrm{i}}{\hbar}\biggl\{(q_{\alpha +}'+q_{\alpha -}')
\int_0^t\df{s}\sum_{k=1}^N\frac{c_{\alpha ,k}^2}{2m_{\alpha,k}\omega_{\alpha ,k}^2}\cos(\omega_{\alpha ,k}s)
\left[q_{\alpha +}(s)-q_{\alpha -}(s)\right]\biggr.\biggr. 
\notag \\
&\phantom{f}+ \biggl.\biggl.\int_0^t\df{s}\int_0^s\df{u}\sum_{k=1}^N\frac{c_{\alpha ,k}^2}{2m_{\alpha,k}
\omega_{\alpha ,k}^2}\cos[\omega_{\alpha ,k}(s-u)]
\left[\dot{q}_{\alpha +}(u)+\dot{q}_{\alpha-}(u)\right]\biggr.\biggr. 
\notag \\
&\phantom{f} \times \biggl.\biggl.\left[q_{\alpha +}(s)-q_{\alpha -}(s)\right]\biggr\}\biggr)
\times \exp\biggl\{-\frac{1}{\hbar}\int_0^t\df{s}\int_0^s\df{u}
\sum_{k=1}^N\frac{c_{\alpha,k}^2}{2m_{\alpha,k}\omega_{
\alpha,k}}
\coth\left(\frac{\hbar\beta_\alpha\omega_{\alpha,k}}{2}\right)\biggr. 
\notag \\
&\phantom{f} \times\biggl.\cos[\omega_{\alpha,k}(u-s)]
\left[q_{\alpha +}(s)-q_{\alpha -}(s)\right]\left[q_{\alpha +}(u)-q_{\alpha -}(u)\right]\biggr\}.
\end{align}

As it is customary in the literature, the limit to the continuum in the spectrum of the bath
is performed by means of the spectral density $J_\alpha(\omega_\alpha)$, which is defined as
\bea
\label{ap-13}
J_\alpha(\omega_\alpha)=\pi\sum_{k=1}^N\frac{c_{\alpha,k}^2}{2m_{\alpha,k}\omega_{\alpha,k}}
\delta(\omega_\alpha-\omega_{\alpha,k}).
\eea
Therefore, for a discrete set modes in the thermal baths, the above spectral density is 
made up of Dirac deltas.
However, for the set of oscillators in the thermal baths to behave as a formal thermal bath, 
it is assumed that the frequencies $\omega_{\alpha,k}$ are so dense as to form a continuous 
spectrum.
Thus, in the continuous limit, the spectral density can be represented as a 
continuous and smooth function on the frequency $\omega_\alpha$\cite{Wei08}. 
This allows for defining the dissipation $\gamma_\alpha(s)$ and noise $K_\alpha(s)$ 
kernels in terms of the spectral density as
\begin{align}
\gamma_\alpha(s) &= \frac{2}{m_\alpha}\int_0^\infty\frac{\df{\omega_\alpha}}{\pi}
\frac{J_\alpha(\omega_\alpha)}{\omega_\alpha}\cos(\omega_\alpha s), 
\label{ap-14} 
\\
K_\alpha(s) &=
\int_0^\infty\frac{\df{\omega_\alpha}}{\pi}J_\alpha(\omega_\alpha)\coth\left(\frac{\hbar\beta_\alpha\omega_\alpha}{
2}\right)\cos(\omega_\alpha s). \label{ap-15}
\end{align}
By replacing (\ref{ap-13}) in the last expressions, 
\begin{align}
\gamma_\alpha(s) &=
\frac{1}{m_\alpha}\sum_{k=1}^N\frac{c_{\alpha,k}^2}{2m_{\alpha,k}\omega_{\alpha,k}^2}\cos(\omega_{\alpha,k}s), 
\label{ap-16}
\\
K_\alpha(s) &=
\sum_{k=1}^N\frac{c_{\alpha,k}^2}{2m_{\alpha,k}\omega_{\alpha,k}}\coth\left(\frac{\hbar\beta_\alpha\omega_{\alpha,k
}}{2}\right)\cos(\omega_{\alpha,k}s), \label{ap-17}
\end{align}
which can be identified with the summations in the argument of the exponentials in Eq.~(\ref{ap-11}).
Having in mind these considerations, and using the expressions (\ref{ap-13}), 
(\ref{ap-14}) and (\ref{ap-15}), Eq.~(\ref{ap-11}) can be expressed in terms of the dissipation 
and noise kernels as
\begin{align}
\label{ap-18}
&\mc{F}[q_{\alpha +},q_{\alpha -}] 
\notag \\
&\phantom{f(a)} =\exp\left(-\frac{\mathrm{i}}{\hbar}\frac{m_\alpha}{2}
\left\{(q_{\alpha +}' +q_{\alpha-}')
\int_0^t\df{s}\gamma_\alpha(s)\left[q_{\alpha +}(s)-q_{\alpha-}(s)\right]\right.\right. 
\notag \\
&\phantom{f(a)} +\left.\left.
\int_0^t\df{s}\int_0^s\df{u}\gamma_\alpha(s-u)\left[\dot{q}_{\alpha +}(u)+\dot{q}_{\alpha
-}(u)\right]
\left[q_{\alpha +}(s)-q_{\alpha -}(u)\right]\right\}\right) 
\notag \\
&\phantom{f(a)} \times \exp\left\{-\frac{1}{\hbar}
\int_0^t\df{s}\int_0^s\df{u}K_\alpha(u-s)\left[q_{\alpha +}(s)-q_{\alpha-}(s)\right]
\left[q_{\alpha +}(u)-q_{\alpha -}(u)\right]\right\}.
\end{align}
By replacing the above expression in (\ref{ap-2}), the total influence functional 
for the system of interest reads
\begin{align}\label{ap-19}
&\mc{F}[q_{1+},q_{2+},q_{1-},q_{2-}] 
\notag \\
&\phantom{f(a)} =\prod_{\alpha=1}^2 \exp\biggl(-\frac{\mathrm{i}}{\hbar}\frac{m_\alpha}{2}
\biggl\{(q_{\alpha +}'+q_{\alpha -}')
\int_0^t\df{s}\gamma_\alpha(s)\left[q_{\alpha +}(s)-q_{\alpha-}(s)\right]\biggr.\biggr. 
\notag \\
&\phantom{f(a)} + \biggl.\biggl.
\int_0^t\df{s}\int_0^s\df{u}\gamma_\alpha(s-u)\left[\dot{q}_{\alpha +}(u)+
\dot{q}_{\alpha-}(u)\right]
\left[q_{\alpha +}(s)-q_{\alpha -}(u)\right]\biggr\}\biggr) 
\notag \\
&\phantom{f(a)} \times \exp\biggl\{-\frac{1}{\hbar}
\int_0^t\df{s}\int_0^s\df{u}K_\alpha(u-s)\left[q_{\alpha +}(s)
- q_{\alpha-}(s)\right]\left[q_{\alpha +}(u)-q_{\alpha -}(u)\right]\biggr\}.
\end{align}

\subsection{Explicit Expression for the Propagating Function}
To provide an explicit expression for the propagating function associated with the system of 
interest, consider the expression (\ref{eq-20})
\begin{align}
\label{ap-20}
&J(q_{1+}'',q_{2+}'',q_{1-}'',q_{2-}'',t;q_{1+}',q_{2+}',q_{1-}',q_{2-}',0) = 
\notag \\
&\phantom{f(a+b)}\int\dfp{q_{1+}}\dfp{q_{2+}}\dfp{q_{1-}}\dfp{q_{2-}}
\exp\biggl\{\frac{\mathrm{i}}{\hbar}\left(S_\mathrm{S}[q_{1+},q_{2+}]-S_\mathrm{S}[q_{1-},q_{2-}]\right)\biggr\} 
\notag \\
&\phantom{f(a+b)} \times \mc{F}[q_{1+},q_{2+},q_{1-},q_{2-}],
\end{align}
with $\mc{F}[q_{1+},q_{2+},q_{1-},q_{2-}]$ given by the expression (\ref{ap-19}). 
Since the path integrals in (\ref{ap-20}) are quadratic, they can be performed exactly\cite{FH05}.
This leaves for the propagating function \cite{ZH95,GPZ10,PR08,GSI88} the following 
expression
\begin{align}
\label{ap-21}
&J(q_{1+}'',q_{2+}'',q_{1-}'',q_{2-}'',t;q_{1+}',q_{2+}',q_{1-}',q_{2-}',0) = 
\notag \\
&\phantom{f(a+b+c+d)}\frac{1}{N(t)}
\exp\left\{\frac{\mathrm{i}}{\hbar}\left(S_\mathrm{S}[\bar{q}_{1+},\bar{q}_{2+}]-
S_\mathrm{S}[\bar{q}_{1-},\bar{q}_{2-}]\right)\right\}\mc{F}[\bar{q}_{1+},\bar{q}_{2+},\bar{q}_{1-},\bar{q}_{2-}],
\end{align}
where $\bar{q}_{\alpha\pm}$ denotes the classical paths of the system and $N(t)$ is a 
normalization factor so that $\mathrm{tr} \hat{\rho}_{\mathrm{S}}(t) = 1$.
Taking into account expressions (\ref{eq-11}) to (\ref{eq-14}) and (\ref{ap-19}), 
the propagating function takes the form
\begin{align}\label{ap-22}
&J(q_{1+}'',q_{2+}'',q_{1-}'',q_{2-}'',t;q_{1+}',q_{2+}',q_{1-}',q_{2-}',0) = 
\notag \\
&\phantom{f} \frac{1}{N(t)}\exp\left\{\frac{\mathrm{i}}{\hbar}
\int_0^t\df{s}
\left[\sum_{\alpha=1}^2\left(\frac{1}{2}m_\alpha\dot{\bar{q}}_{\alpha +}^2(s)
-\frac{1}{2}m_\alpha\omega_\alpha^2\bar{q}_{\alpha +}^2(s)\right)-
c(s)\bar{q}_{1 +}(s)\bar{q}_{2+}(s)\right]\right\} 
\notag \\
&\phantom{f} \times \exp\left\{-\frac{\mathrm{i}}{\hbar}
\int_0^t\df{s} \left[\sum_{\alpha=1}^2\left(\frac{1}{2}m_\alpha\dot{\bar{q}}_{\alpha -}^2(s)
-\frac{1}{2}m_\alpha\omega_\alpha^2\bar{q}_{\alpha -}^2(s)\right)-
c(s)\bar{q}_{1 -}(s)\bar{q}_{2-}(s)\right]\right\}
 \notag \\
&\phantom{f} \times \mc{F}[\bar{q}_{1+},\bar{q}_{2+},\bar{q}_{1-},\bar{q}_{2-}].
\end{align}
To simplify the last expression, consider a new set of coordinates\cite{GPZ10,GSI88,ZH95}
\bea\label{ap-23}
Q_\alpha=\frac{1}{2}(\bar{q}_{\alpha +}+\bar{q}_{\alpha -}), 
\quad 
q_\alpha=\bar{q}_{\alpha +}-\bar{q}_{\alpha -},
\eea
where the Jacobian of the transformation is equal to one. 
Hence, Eq.~(\ref{ap-22}) takes the form
\begin{align}
\label{ap-24}
&J(\mathbf{Q}'',\mathbf{q}'',t;\mathbf{Q}',\mathbf{q}',0)  = \notag \\
&\phantom{f} \frac{1}{N(t)} \exp\left(\frac{\mathrm{i}}{\hbar}
\int_0^t\df{s}\left\{
\sum_{\alpha=1}^2\left[m_\alpha\dot{Q}_\alpha(s)\dot{q}_\alpha(s)-m_\alpha
\omega_\alpha^2Q_\alpha(s)q_\alpha(s)
-m_\alpha Q_\alpha'\gamma_\alpha(s)q_\alpha(s)\right.\right.\right. 
\notag \\
&\phantom{f} \left.\left.\left.-\int_0^s \df{u}\gamma_\alpha(s-u)\dot{Q}_\alpha(u)q_\alpha(s)\right]-
c(s)\left[Q_1(s)q_2(s)+Q_2(s)q_1(s)\right]\right\}\right) 
\notag \\
&\phantom{f} \times \exp\left\{-\frac{1}{\hbar}\int_0^t\df{s}\int_0^s\df{u}
\sum_{\alpha=1}^2K_\alpha(u-s)q_\alpha(s)q_\alpha(u)\right\},
\end{align}
where $\mathbf{Q} = (Q_1,Q_2)$ and $\mathbf{q}=(q_1,q_2)$.
The next step is to determine and solve the equations of motion for the coordinates 
$Q_\alpha$ and $q_\alpha$. 
For this, it is necessary to use the Euler-Lagrange equations for the Lagrangian $L$ 
that appears as the integrand in the imaginary phase of the exponential in
Eq.~(\ref{ap-24}).
For the coordinates $Q_\alpha$, the equations of motion have the following form
\bea
\label{ap-25}
\frac{\mathrm{d}}{\mathrm{d}s}
\left(\frac{\partial L}{\partial \dot{q}_\alpha}\right)-\frac{\partial L}{\partial q_\alpha}=0,
\eea
whereas for the coordinates $q_\alpha$, the Euler-Lagrange equations are
\bea
\label{ap-26}
\frac{\mathrm{d}}{\mathrm{d}s}
\left(\frac{\partial L}{\partial \dot{Q}_\alpha}\right)-\frac{\partial L}{\partial Q_\alpha}=0.
\eea
By using Eq.~(\ref{ap-25}) in (\ref{ap-24}), the equation of motion for the coordinate 
$Q_\alpha$ can be written as
\begin{align}
\label{ap-27}
\begin{split}
\ddot{Q}_1(s)+\omega_1^2Q_1(s)+\frac{c(s)}{m_1}Q_2(s)
+\frac{\mathrm{d}}{\mathrm{d}s}\int_0^s\df{u}\gamma_1(s-u)Q_1(u) &= 0, 
\\
\ddot{Q}_2(s)+\omega_2^2Q_2(s)+\frac{c(s)}{m_2}Q_1(s)
+\frac{\mathrm{d}}{\mathrm{d}s}\int_0^s\df{u}\gamma_2(s-u)Q_2(u) &= 0.
\end{split}
\end{align}
In order to find the equations of motion for the coordinates $q_\alpha$, consider the 
following equality
\begin{multline}\label{ap-28}
Q_\alpha'\int_0^t\df{s}\gamma_\alpha(s)q_\alpha(s)
+\int_0^t\df{s}\int_0^s\df{u}\gamma_\alpha(s-u)\dot{Q}_\alpha(u)q_\alpha(s) = 
\\
q_\alpha''\int_0^t\df{s}\gamma_\alpha(t-s)Q_\alpha(s)
-\int_0^t\df{s}\int_s^t\df{u}\gamma_\alpha(u-s)\dot{q}_\alpha(u)Q_\alpha(s).
\end{multline}
Thus, the propagating function in Eq.~(\ref{ap-24}) can be written as
\begin{align}
\label{ap-29}
&J(\mathbf{Q}'',\mathbf{q}'',t;\mathbf{Q}',\mathbf{q}',0)  = 
\notag \\
&\phantom{f(a+b+c)} \frac{1}{N(t)} \exp\biggl(\frac{\mathrm{i}}{\hbar}
\int_0^t\df{s}\biggl\{\sum_{\alpha=1}^2
\biggl[m_\alpha\dot{Q}_\alpha(s)\dot{q}_\alpha(s)-m_\alpha(s)\omega_\alpha^2
Q_\alpha(s)q_\alpha(s)\biggr.\biggr.\biggr. 
\notag \\
&\phantom{f(a+b+c)} \biggl.-q_\alpha''\gamma_\alpha(t-s)Q_\alpha(s)
+\int_s^t\df{u}\gamma_\alpha(u-s)\dot{q}_\alpha(u)Q_\alpha(s)\biggr] 
\notag \\
&\phantom{f(a+b+c)} \biggl.\biggl.-c(s)\left[Q_1(s)q_2(s)+Q_2(s)q_1(s)\right]\biggr\}\biggr) 
\notag \\
&\phantom{f(a+b+c)} \times \exp\left\{-\frac{1}{\hbar}\int_0^t\df{s}
\int_0^s\df{u}\sum_{\alpha=1}^2 K_\alpha(u-s)q_\alpha(s)q_\alpha(u)\right\}.
\end{align}
By virtue of the expression (\ref{ap-26}), the equations of motion for the coordinates $q_\alpha$
are given by
\begin{align}
\label{ap-30}
\begin{split}
\ddot{q}_1(s)+\omega_1^2q_1(s)
+\frac{c(s)}{m_1}q_2(s)-\frac{\mathrm{d}}{\mathrm{d}s}\int_s^t\df{u}\gamma_1(u-s)q_1(u) &=0, 
\\
\ddot{q}_2(s)+\omega_2^2q_2(s)+\frac{c(s)}{m_2}q_1(s)
-\frac{\mathrm{d}}{\mathrm{d}s}\int_s^t\df{u}\gamma_2(u-s)q_2(u) &=0.
\end{split}
\end{align}
The coordinates $Q_\alpha$ and $q_\alpha$ satisfy the boundary conditions
\bea
\label{ap-31}
Q_\alpha(s)=
\begin{cases} Q_\alpha', & s=0, 
\\ 
Q_\alpha'', & s=t,\end{cases} 
\quad 
q_\alpha(s)=\begin{cases} q_\alpha', & s=0, 
\\ 
q_\alpha'', & s=t. 
\end{cases}
\eea
Since the equations of motion (\ref{ap-28}) and (\ref{ap-31}) are linear, 
the solution to them can be written as in Eq.~(\ref{eq:SltnsExtrmCndtns}).
Once the solutions for the equations of motion are calculated, only one step is left to find 
the propagating function. 
The partial integration of the first term in the imaginary phase in the propagating 
function in (\ref{ap-24}) leads to
\begin{align*}
&J(\mathbf{Q}'',\mathbf{q}'',t;\mathbf{Q}',\mathbf{q}',0)  = 
\\
&\phantom{f(a+b+c)} \frac{1}{N(t)} 
\exp\biggl(\frac{\mathrm{i}}{\hbar}\biggl\{q_1''\dot{Q}_1(t)-q_1'\dot{Q}_1(0)
+q_2''\dot{Q}_2(t)-q_2'\dot{Q}_2(0)\biggr.\biggr. \\
&\phantom{f(a+b+c)} -m_1\int_0^t\df{s}\biggl[\ddot{Q}_1(s)+\omega_1^2
Q_1(s)+\frac{c(s)}{m_1}Q_2(s)+Q_1'\gamma_1(s) \biggr.
\notag \\
&\phantom{f(a+b+c)} +\biggl.\int_0^s\df{u}\gamma_1(s-u)\dot{Q}_1(u)\biggr]q_1(s)
-m_2\int_0^t\df{s}\biggl[\ddot{Q}_2(s)+\omega_2^2Q_2(s) \biggr.
 \\
&\phantom{f(a+b+c)} \biggl.+\frac{c(s)}{m_2}Q_1(s)+Q_2'\gamma_2(s)
+\int_0^s\df{u}\gamma_2(s-u)\dot{Q}_2(u)\biggr]q_2(s)\biggr\}\biggr) 
\\
&\phantom{f(a+b+c)} \times \exp\left\{-\frac{1}{\hbar}\int_0^t\df{s}
\int_0^s\df{u}\sum_{\alpha=1}^2 K_\alpha(u-s)q_\alpha(s)q_\alpha(u)\right\}.
\end{align*}
The last two terms inside the integrals in the imaginary phase of the exponential 
are the classical equations of motion in (\ref{ap-28}) and (\ref{ap-30}), therefore, 
their contribution vanish.
Hence, the propagating function takes the compact form
\begin{align}
J(\mathbf{Q}'',\mathbf{q}'',t;\mathbf{Q}',\mathbf{q}',0) &= 
 \frac{1}{N(t)} 
\exp\biggl\{\frac{\mathrm{i}}{\hbar}\sum_{\alpha=1}^2\left[q_\alpha''\dot{Q}_\alpha(t)
-q_\alpha'\dot{Q}_\alpha(0)\right]\biggr. 
\\
& \biggl.-\frac{1}{\hbar}\int_0^t\df{s}
\int_0^s \df{u}\sum_{\alpha=1}^2K_\alpha(u-s)q_\alpha(s)q_\alpha(u)\biggr\}.
\end{align}
By replacing Eq.~(\ref{eq:SltnsExtrmCndtns}) into the last expression, the propagating function 
reads
\begin{align}
\label{ap-33}
&J(\mathbf{Q}'',\mathbf{q}'',t;\mathbf{Q}',\mathbf{q}',0) = 
\notag \\
&\phantom{f(a+b)} \frac{1}{N(t)} \exp\biggl\{\frac{\mathrm{i}}{\hbar}
\biggl[q_1''\left(\dot{U}_1(t,t)Q_1'+\dot{U}_2(t,t)Q_1''+\dot{U}_3(t,t)Q_2'
+\dot{U}_4(t,t)Q_2''\right)\biggr.\biggr. 
\notag \\
&\phantom{f(a+b)} -q_1'\left(\dot{U}_1(t,0)Q_1'+\dot{U}_2(t,t)Q_1''+\dot{U}_3(t,0)Q_2'
+\dot{U}_4(t,0)Q_2''\right) 
\notag \\
&\phantom{f(a+b)} +q_2''\left(\dot{V}_1(t,t)Q_2'+\dot{V}_2(t,t)Q_2''+\dot{V}_3(t,t)Q_1'
+\dot{V}_4(t,t)Q_1''\right) 
\notag \\
&\phantom{f(a+b)} \biggl.\biggl.-q_2'\left(\dot{V}_1(t,0)Q_2'+\dot{V}_2(t,0)Q_2''
+\dot{V}_3(t,0)Q_1'+\dot{V}_4(t,0)Q_1''\right)\biggr]\biggr\} 
\notag \\
&\phantom{f(a+b)} \times\exp\biggl\{-\frac{1}{\hbar}\int_0^t\df{s}\int_0^s\df{u}
K_1(u-s)\left[u_1(t,s)q_1'+u_2(t,s)q_1''+u_3(t,s)q_2'\right. 
\notag \\
&\phantom{f(a+b)} \biggl.\left.+u_4(t,s)q_2''\right]\times
\left[u_1(t,u)q_1'+u_2(t,u)q_1''+u_3(t,u)q_2'+u_4(t,u)q_2''\right]\biggr\} 
\notag \\
&\phantom{f(a+b)} \times\exp\biggl\{-\frac{1}{\hbar}\int_0^t\df{s}\int_0^s\df{u}
K_2(u-s)\left[v_1(t,s)q_2'+v_2(t,s)q_2''+v_3(t,s)q_1'\right.\biggr. 
\notag \\
&\phantom{f(a+b)} \biggl.\left.+v_4(t,s)q_1''\right]\times
\left[v_1(t,u)q_2'+v_2(t,u)q_2''+v_3(t,u)q_1'+v_4(t,u)q_1''\right]\biggr\}.
\end{align}
Or alternatively,
\bea
\label{ap-34}
J(\mathbf{Q}'',\mathbf{q}'',t;\mathbf{Q}',\mathbf{q}',0)=\frac{1}{N(t)}
\exp\left\{\frac{\mathrm{i}}{\hbar}\vtr{x}_1\cdot{\mathsf{A}}\cdot\vtr{x}_1
-\frac{1}{\hbar}\vtr{x}_2\cdot\mathsf{B}\cdot\vtr{x}_2\right\},
\eea
where $\vtr{x}_1 = (\mathbf{Q}'',\mathbf{Q}',\mathbf{q}'',\mathbf{q}')$,
$\vtr{x}_2 = (\mathbf{q}'',\mathbf{q}')$ and the matrix elements $A_{ij}$
and $B_{ij}$ of the matrices $\mathsf{A}$ and $\mathsf{B}$ are given by
\begin{align}\label{ap-35}
\begin{split}
A_{ij} &= \frac{1}{2}\frac{\partial}{\partial x_1^i}
\frac{\partial \phi_\mathrm{I}}{\partial x_1^j}, 
\qquad
B_{ij} = \frac{1}{2}\frac{\partial}{\partial x_2^i}
\frac{\partial \phi_\mathrm{R}}{\partial x_2^j}, 
\end{split}
\end{align}
respectively. 
Here $\phi_\mathrm{I}$ and $\phi_\mathrm{R}$ are the imaginary and real components,
respectively, of the phase in the propagating function.
In terms of the matrix elements $A_{ij}$ and $B_{ij}$, the propagating function can be written as
\begin{align}
\label{ap-36}
J(\mathbf{Q}'',\mathbf{q}'',t;\mathbf{Q}',\mathbf{q}',0) &= 
\frac{1}{N(t)}\exp\biggl(\frac{2\mathrm{i}}{\hbar}
\biggl\{q_1''\biggl[A_{15}(t)Q_1''+A_{25}(t)Q_2''+A_{35}(t)Q_1'+A_{45}(t)Q_2'\biggr]\biggr.\biggr. 
\notag \\
&+ q_2''\biggl[A_{16}(t)Q_1''+A_{26}(t)Q_2''+A_{36}(t)Q_1'+A_{46}(t)Q_2'\biggr] 
\notag \\
&- q_1'\biggl[A_{17}(t)Q_1''+A_{27}(t)Q_2''+A_{37}(t)Q_1'+A_{47}(t)Q_2'\biggr] 
\notag \\
&- \biggl.\biggl.q_2'\biggl[A_{18}(t)Q_1''+A_{28}(t)Q_2''+A_{38}(t)Q_1'+A_{48}(t)Q_2'\biggl]\biggr\}\biggr) \\
&\times \exp\biggl\{-\frac{1}{\hbar}\biggl[q_1''^2 B_{11}(t)+q_2''^2B_{22}(t)+q_1'^2B_{33}(t)+q_2'^2B_{44}(t)\biggr.\biggr. \notag \\
&+ 2 q_1''q_2''B_{12}(t)+2 q_1''q_1'B_{13}(t)+2 q_1''q_2'B_{14}(t) \notag \\
&+ \biggl.\biggl.q_1'q_2''B_{23}(t)+2q_2''q_2'B_{24}(t)+2q_1'q_2'B_{34}(t)\biggr]\biggr\}.
\end{align}

\end{document}